\documentclass[aps,preprint,prd,nofootinbib]{revtex4-1}
\pdfoutput=1
\usepackage{amsmath}
\usepackage{xfrac}
\usepackage[colorlinks, linkcolor=black, anchorcolor=black, citecolor=black]{hyperref}
\usepackage{graphicx}
\usepackage{psfrag}
\usepackage{float}
\usepackage{subfigure}
\usepackage{array}
\usepackage{amssymb}
\usepackage{mathrsfs}
\usepackage{bm}
\usepackage{slashed}
\usepackage{threeparttable}
\usepackage{multirow}
\usepackage[amsmath,thmmarks,hyperref]{ntheorem}
\usepackage{soul}
\allowdisplaybreaks[4]

\begin{document}

\title{The study of two quasi-degenerate heavy sterile neutrinos in rare meson decays}
\author{Jiabao Zhang$^1$\footnote{19s011001@stu.hit.edu.cn}, Tianhong Wang$^1$\footnote{thwang@hit.edu.cn (Corresponding author)}, Geng Li$^1$\footnote{karlisle@hit.edu.cn}, Yue Jiang$^1$\footnote{jiangure@hit.edu.cn}, Guo-Li Wang$^{1, 2, 3}$\footnote{gl\_wang@hit.edu.cn}\\}
\address{$^1$School of Physics, Harbin Institute of Technology, Harbin, 150001, China\\
$^2$Department of Physics, Hebei University, Baoding 071002, China\\
$^3$Hebei Key Laboratory of High-precision Computation and Application of Quantum Field Theory, Baoding
071002, China}

\baselineskip=20pt

\begin{abstract}

In this work, we study the lepton-number-violating processes of $K^\pm$ and $D^\pm$ mesons. Two quasi-degenerate sterile neutrinos are assumed to induce such processes. Different with the case where only one sterile neutrino involves, here, the CP phases of the mixing parameters could give sizable contribution. This, in turn, would affect the absolute values of the mixing parameters determined by the experimental upper limits of the branching fractions. A general function which express the difference of the mixing parameters for two-generation and one-generation is presented. Special cases with specific relations of the parameters are discussed. Besides, we also thoroughly investigate the CP violation effect of such processes. It is shown that generally $\mathcal A_{CP}$ is a function of the sterile neutrino mass.

\end{abstract}

\maketitle

\section{Introduction}

The neutrino oscillation phenomena indicate that at least two of the three active neutrinos have nonzero masses. Usually, different see-saw mechanisms are proposed to explain why the masses of active neutrinos are so small. For example, in the type-I see-saw mechanism, the right-handed sterile neutrinos which are in the GUT scale are introduced. However, there are also models which allow the existence of  keV or GeV sterile neutrinos~\cite{Asaka:2005an,Asaka:2005pn}. Neutrinos at these mass scales could be produced on-shell in the meson rare decays, which can be studied at B-factories. A clear signal which indicates the sterile neutrinos being of Majorana type is the observation of the lepton-number-violating (LNV) processes of charged mesons, which has been extensively studied theoretically in Refs.~\cite{Atre:2005eb,Dib:2000wm,Ali:2001gsa,Zhang:2010um,Yuan:2013yba} by assuming one extra sterile neutrino $N_4$. Besides, the $N_4$ induced LNV processes of tau lepton~\cite{Ilakovac:1995km,Ilakovac:1994kj,Ilakovac:1995wc,Gribanov:2001vv,Atre:2005eb,Godbole:2020doo,Yuan:2017uyq} and baryons~\cite{Mejia-Guisao:2017nzx} have also been investigated. By comparing with the experimental data of the branching ratios, the upper limits of the mixing parameters $U_{\ell4}$ can be obtained.

If two extra sterile neutrinos $N_4$ and $N_5$ are introduced, some new aspects should be considered. In Ref.~\cite{Abada:2019bac}, Abada {\sl et al} have shown that when mediated by two generations of quasi-degenerate neutrinos, interference effect will make the observation of LNV process and lepton-flavor-violating (LFV) process of semileptonic meson decays complement each other, and the non-observation of  LNV process in current experiments does not necessarily lead to more stringent bounds on the corresponding mixing matrix elements.  In Refs.~\cite{Cvetic:2013eza,Cvetic:2014nla,Cvetic:2015naa,Dib:2014pga}, the CP asymmetry in rare meson decays caused by the interference between two quasi-degenerate generations of sterile Majorana neutrino have also been extensively studied. In Ref.~\cite{Zamora-Saa:2016ito}, the resonant CP violation in rare $\tau^\pm$ decays has also been considered.

However, there are still two things about such decays deserve further studies. Firstly, with two quasi-degenerate sterile neutrinos, the CP phase  may play an important role, especially in some specific parameter space. This will affect the determination of the upper limits of the mixing parameters. For example, there may be large cancelation when the CP phases approach to $\pi$. When we use the experimental values to set the upper limits of $U_{\ell4, 5}$, they could be much larger than those of the one-generation case. And when we use such limits to set the upper bounds of the branching ratios of other meson decays, such as $B$ meson, the results will also be changed. Secondly, the CP asymmetry in such decays will in general depend on several parameters, such as the ratios of $|U_{\ell4, 5}|$, the CP phases, and the sterile neutrino mass. A thorough study of how the CP asymmetry changes with these parameters is necessary. In this paper, we will try to study such two things.

The paper is organized as follows. In Sec.~II, we present the calculation of the decay width of the LNV processes for $K$ meson. A function which is defined as the ratio of $|U_{\ell4}|$ in the two-generation and one-generation cases are obtained. In Sec.~III, two special cases are considered. In Sec.~IV, we investigate the CP violation effect, and discuss how the CP asymmetry changes with related parameters. Finally, we present the conclusion in Sec.~V.

\section{General Consideration}

If only one-generation of heavy sterile neutrino is assumed, the decay width will just depend on $|U_{\ell4}|$, while the CP phases have no influence on the physical results. This situation has been extensively studied, such as in Refs.~\cite{Atre:2009rg,Abada:2017jjx}. However, if there exist two generations of sterile neutrinos, the CP phases will be relevant. As in Ref.~\cite{Abada:2019bac}, we will parameterize  the active-sterile mixing matrix elements as $U_{\ell N}=|U_{\ell N}|e^{-i\phi_{\ell N}}$, where $\ell=e,\mu,\tau, N=4,5$ and the CP phase $\phi_{\ell N}$ contains both the Dirac and Majorana phases. 

The decay width of the heavy sterile neutrino can be written as
\begin{equation}\label{g4split}
\Gamma_N=|U_{eN}|^2f_e(m_N)+|U_{\mu N}|^2f_\mu(m_N)+|U_{\tau N}|^2f_\tau(m_N),
\end{equation}
where $f_\ell$ ($\ell=e,\mu,\tau$), as functions of $m_N$, are achieved by considering all the possible decay channels of the sterile neutrino (see Ref.~\cite{Atre:2009rg}). As an example, we set $|U_{lN}|$ to one and plot $\Gamma_N$ and $f_\ell/\Gamma_N$ in Fig.~1(a) and (b), respectively. Generally, two generations of sterile neutrinos may have different widths because of different mixing parameters and masses. We can write their ratio as,
\begin{equation}\label{k}
\begin{aligned}
k&=\frac{|U_{e5}|^2f_e(m_5)+|U_{\mu 5}|^2f_\mu(m_5)+|U_{\tau 5}|^2f_\tau(m_5)}{|U_{e4}|^2f_e(m_4)+|U_{\mu 4}|^2f_\mu(m_4)+|U_{\tau 4}|^2f_\tau(m_4)}\\
&\approx\frac{k_ef_e(m_4)+k_{\mu e}k_{\mu}f_\mu(m_4)+k_{\tau e}k_{\tau}f_\tau(m_4)}{f_e(m_4)+k_{\mu e}f_\mu(m_4)+k_{\tau e}f_\tau(m_4)},
\end{aligned}
\end{equation}
where we have defined 
\begin{equation}\label{ks}
k_e=\frac{|U_{e5}|^2}{|U_{e4}|^2},~k_{\mu}=\frac{|U_{\mu5}|^2}{|U_{\mu4}|^2},~k_{\tau}=\frac{|U_{\tau5}|^2}{|U_{\tau4}|^2},~k_{\mu e}=\frac{|U_{\mu4}|^2}{|U_{e4}|^2},~k_{\tau e}=\frac{|U_{\tau4}|^2}{|U_{e4}|^2}.
\end{equation}
To get the second line of Eq.~(2), we have assumed the Majorana neutrinos being quasi-degenerate, namely $\Delta m\equiv m_5-m_4\ll m_4$. This is reasonable, because from Fig.1 we see $\Gamma_N\ll m_4$, and the interference effect is important only when $\Delta m\sim \Gamma_N$. 

\begin{figure}[htbp]
	\begin{minipage}{0.45\linewidth}
		\centerline{\includegraphics[width=7.25cm]{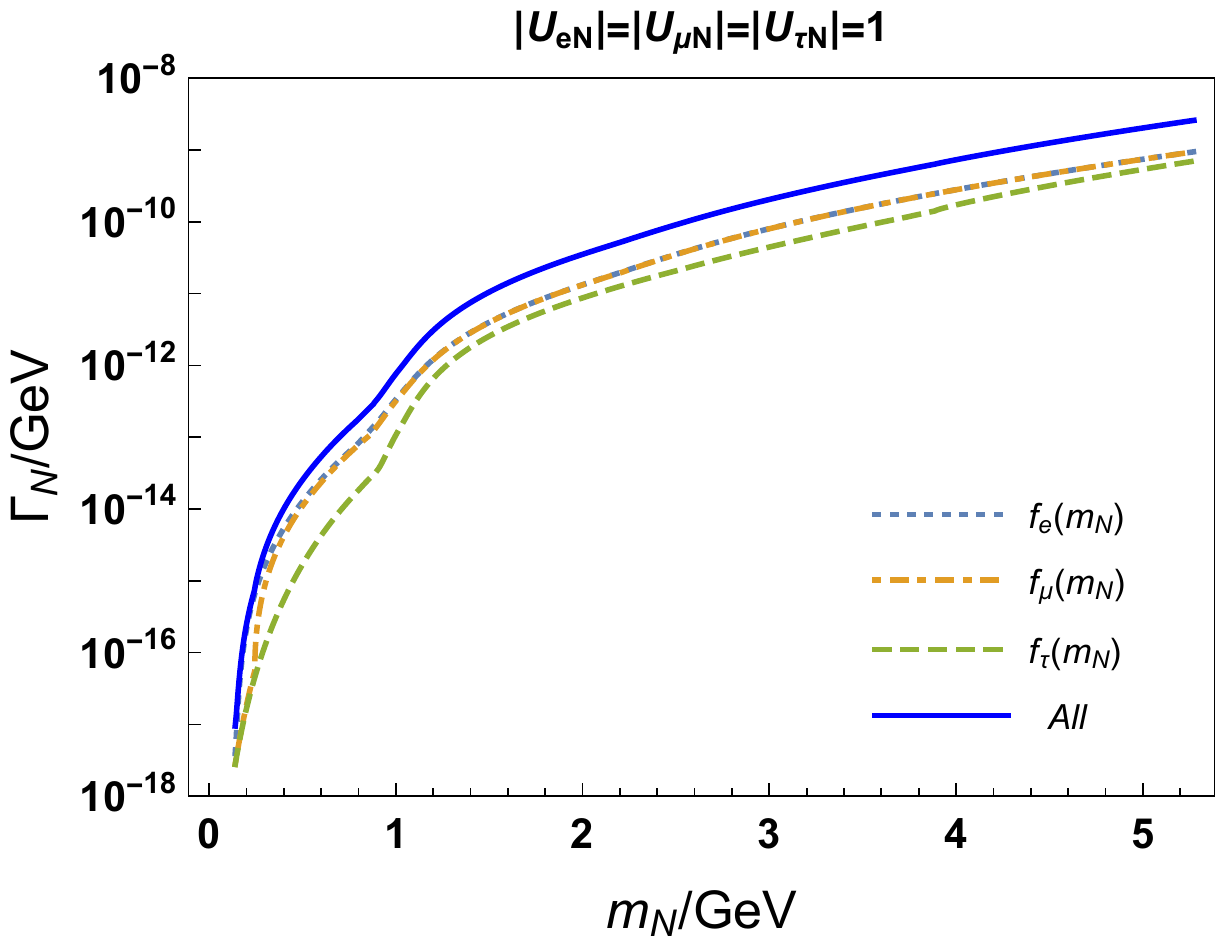}}
		     \centerline{(a)}
	\end{minipage}
	\hfill
	\begin{minipage}{0.45\linewidth}
		\centerline{\includegraphics[width=7.0cm]{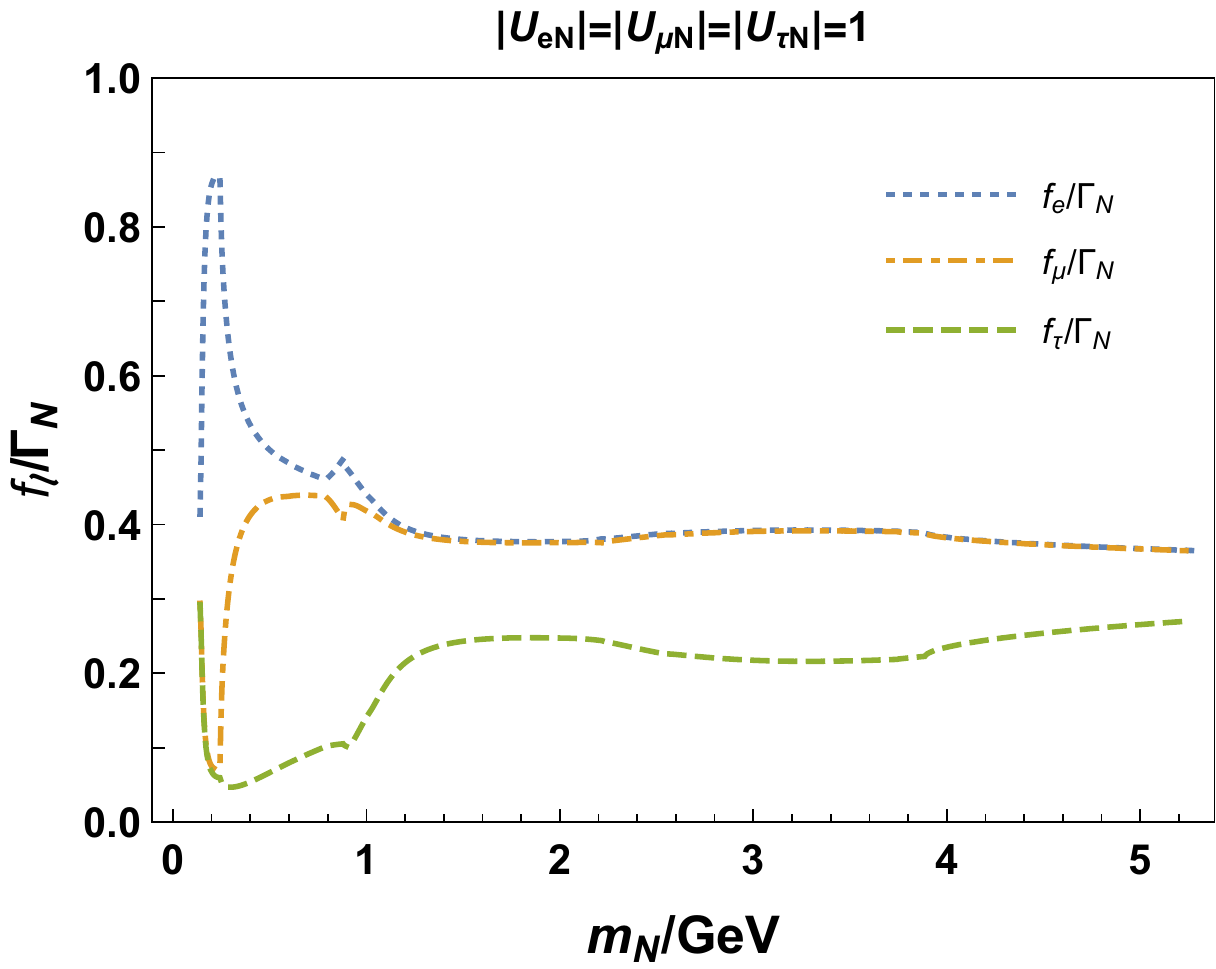}}
		     \centerline{(b)}
	\end{minipage}
	\vfill
	\caption{(a) Decay width of heavy sterile neutrino with $m_N\in$[0.140,5.279]~GeV; (b) the portion of the contribution made by $U_e$, $U_{\mu}$ and $U_{\tau}$, respectively.}
	\label{g4}
\end{figure}

\begin{figure}[htbp]
	\centering
	\includegraphics[width=0.65\textwidth,scale=0.9]{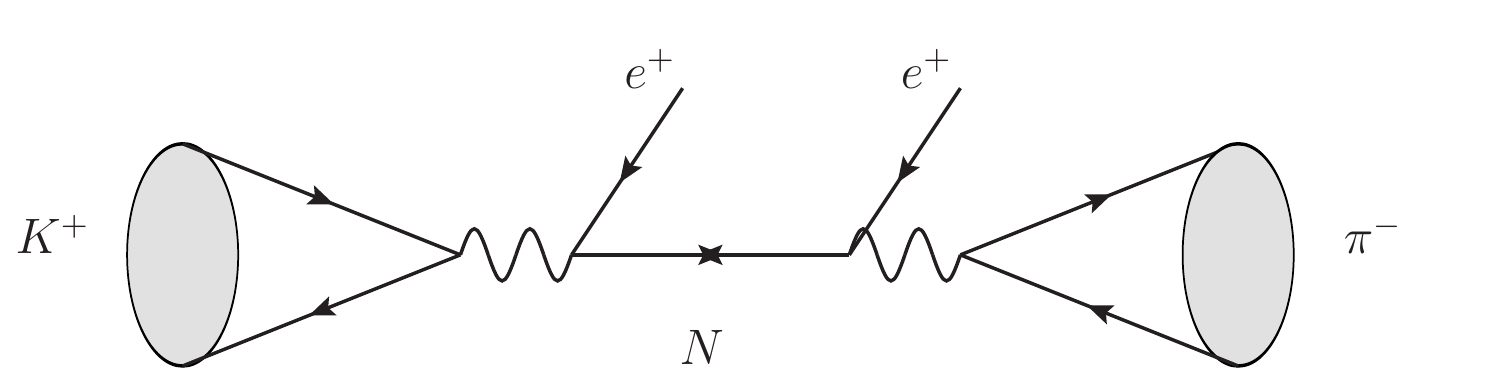}
	\caption{Feynman diagram of the LNV process $K^+\rightarrow e^+e^+\pi^-$.}
	\label{FeynKeepi}
\end{figure}

We will consider the LNV process $K^+(P)\rightarrow e^+(p_1)e^+(p_2)\pi^-(p_3)$, whose  Feynman diagram is shown in Fig.~\ref{FeynKeepi}.
Following the Feynman rules in Ref.~\cite{Atre:2009rg}, we write the amplitude of this process as
\begin{equation}\label{amp2}
\begin{aligned}
i\mathcal{M}&=2G_F^2f_Kf_{\pi}V_{ud}V_{us}\left[\frac{U_{e4}U_{e4}m_4}{s_{23}-m_4^2+i\Gamma_4m_4}+\frac{U_{e5}U_{e5}m_5}{s_{23}-m_5^2+i\Gamma_5m_5}\right]\bar{u}(p_1)\slashed{P}\slashed{p}_3P_Rv(p_2),
\end{aligned}
\end{equation}
where $G_F$ is the Fermi constant; $f_K$ and $f_\pi$ are the decay constants of $K$ and $\pi$, respectively; $s_{23}=(p_2+p_3)^2$. Here we do not include the contribution of the exchange diagram, because $\Gamma_N$ is too small and the interference terms between two diagrams can be neglected. Correspondingly, when doing the phase space integral, one should drop the factor $1/2$. 

We define $x=(s_{23}^2-m_4^2)/(\Gamma_4m_4)$, $x_0=\Delta m/\Gamma_4$, $\Delta\varphi=2(\phi_{e5}-\phi_{e4})$. The square of the absolute value of the $[...]$ part in Eq.~(4) can be written as
\begin{equation}
\left|\frac{U_{e4}U_{e4}m_4}{s_{23}-m_4^2+i\Gamma_4m_4}+\frac{U_{e5}U_{e5}m_5}{s_{23}-m_5^2+i\Gamma_5m_5}\right|^2=\frac{|U_{e4}|^4}{\Gamma_4^2}y(k_e,k,x_0,\Delta\varphi,x),
\end{equation}
where
\begin{equation}
\begin{aligned}
y(k_e,k,x_0,\Delta\varphi,x)&=\frac{1}{1+x^2}\bigg\{1+\frac{k_e^2(x^2+1)}{k^2+(x-2x_0)^2}+\frac{2k_e}{k^2+(x-2x_0)^2}\\
&\quad\times\left[(k + x^2 - 2 x x_0) \cos\Delta\varphi-(k x - x + 2 x_0)\sin\Delta\varphi)\right]\bigg\}.
\end{aligned}
\end{equation}
Then the decay width can be expressed as
\begin{equation}
\Gamma=C_f\frac{|U_{e4}|^4m_4}{\Gamma_4}\int y(k_e,k,x_0,\Delta\varphi,x)\text{ILT}(s_{23})dx,
\end{equation}
where $C_f=G_F^2f_K^2f_\pi^2|V_{ue}V_{us}|^2/(4m\pi)^3$ and the $\text{ILT}$ function is
\begin{equation}
\begin{aligned}
\text{ILT}(s_{23})&=\frac{1}{2s_{23}^2}\sqrt{m^4-2m^2(m_1^2+s_{23})+(m_1^2-s_{23})^2}\sqrt{m_2^4-2m_2^2(m_3^2+s_{23})+(m_3^2-s_{23})^2}\\
&\quad\times\left[m^2(m_1^2+s_{23})-(m_1^2-s_{23})^2\right]\left[m_2^4-m_2^2(m_3^2+2s_{23})-s_{23}(m_3^2-s_{23})\right].
\end{aligned}
\end{equation}

From the definition of $x$, we can see that even $s_{12}$ has a very small variation from $m_4$, $x$ will change a lot, because $\Gamma_4$ is extremely small compared with $m_4$. This means we can set $s_{23}\approx s_{23}(x=0)=m_4^2$ and take the integral interval to be $(-\infty,~\infty)$. Then the decay width will be expressed as
\begin{equation}\label{2gen}
\begin{aligned}
\Gamma&\approx C_f \frac{|U_{e4}|^4m_4}{\Gamma_4}\int_{-\infty}^{\infty}y(k_e,k,x_0,\Delta\varphi,x)\text{ILT}(m_4^2)dx\\
&=C_f \frac{|U_{e4}|^4m_4}{\Gamma_4} \text{Iy}(k_e,k,x_0,\Delta\varphi)\textrm{ILT}(m_4^2),
\end{aligned}
\end{equation}
where
\begin{equation}\label{IyGeneral}
\text{Iy}(k_e,k,x_0,\Delta\varphi)=\pi\left(1+\frac{k_e^2}{k}\right)+\frac{4\pi k_e}{(k+1)^2+4 x_0^2}\left[(k+1) \cos\Delta \varphi-2 x_0\sin\Delta\varphi\right].
\end{equation}

By using the branching ratio $\text{Br}(K^+\rightarrow e^+e^+\pi^-)$ and the life time of $K^+$ $\tau(K^+)$, we can express the mixing parameter $|U_{e4}|^2$ as
\begin{equation}\label{Ulgeneral}
|U_{e4}|^2=\frac{\text{Br}(K^+\rightarrow e^+e^+\pi^-)}{\tau(K^+)}\frac{g_4(k_{\mu},k_{\tau},m_4)}{C_fm_4\text{Iy}(k,x_0,\Delta\varphi)\text{ILT}(m_4^2)},
\end{equation}
where $g_4(k_{\mu},k_{\tau},m_4)\equiv\Gamma_4/|U_{e4}|^2$. For the one-generation case, one just needs to replace the $\text{Iy}$ function in Eq.~\eqref{Ulgeneral} to $\pi$. To compare the results of two situations, we define a ratio function:
\begin{equation}
R_{21} (k_e,k,x_0,\Delta\varphi)\equiv\frac{|U_{e4}|_{2-gen}}{|U_{e4}|_{1-gen}}=\sqrt{\pi/\text{Iy}(k_e,k,x_0,\Delta\varphi)},
\end{equation}
which depends on $k_e$ both directly an indirectly through $k$. In the next section, we will consider some special situations to study the characteristics of this function.

\section{Some Special Cases}

\begin{figure}[htbp]
\begin{minipage}{0.45\linewidth}
	\centerline{\includegraphics[width=7.0cm]{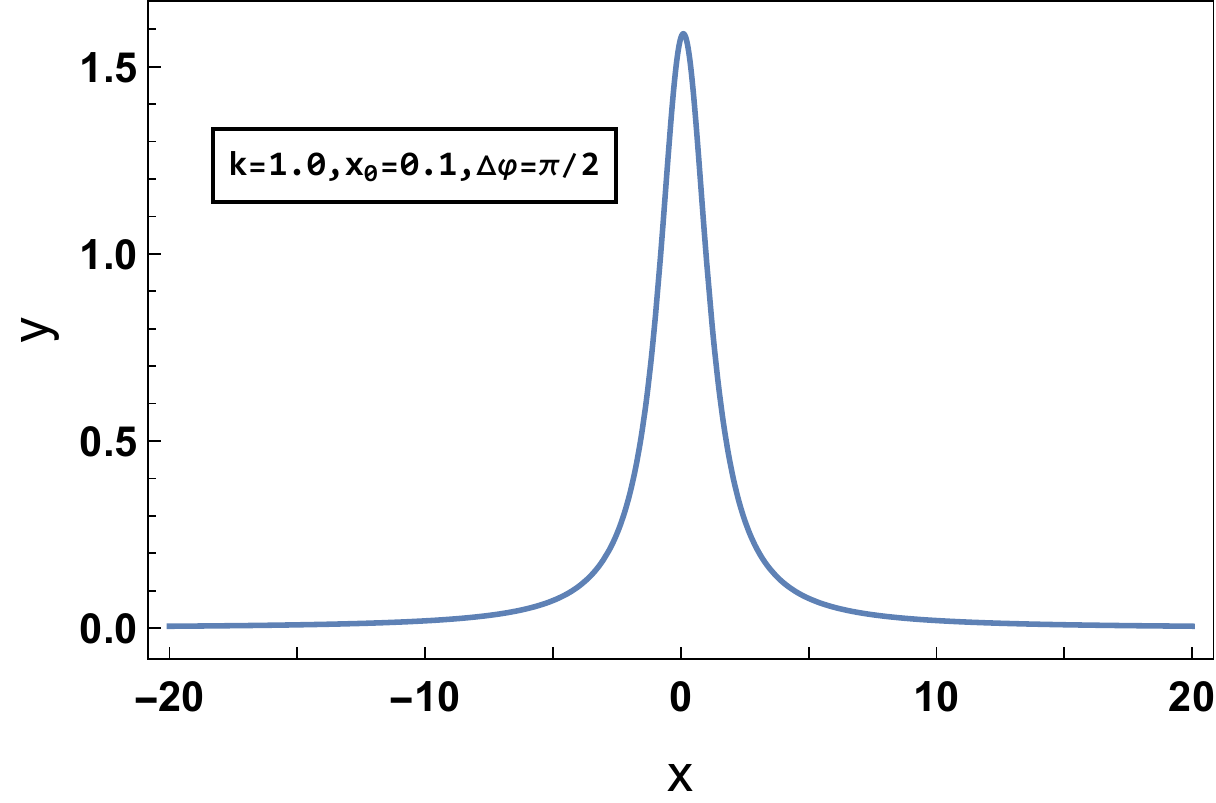}}
     \centerline{(a)}
\end{minipage}
\hfill
\begin{minipage}{0.45\linewidth}
	\centerline{\includegraphics[width=7.0cm]{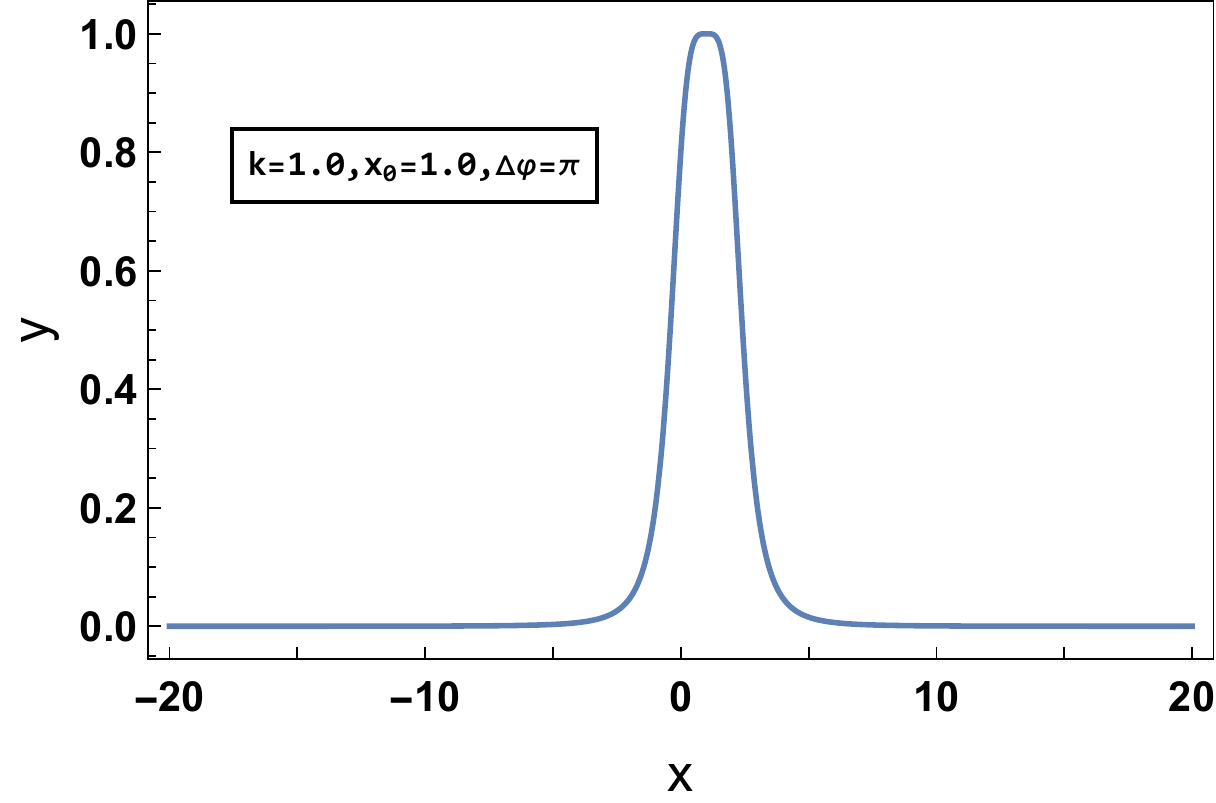}}
     \centerline{(b)}
\end{minipage}
\vfill
\begin{minipage}{0.45\linewidth}
	\centerline{\includegraphics[width=7.0cm]{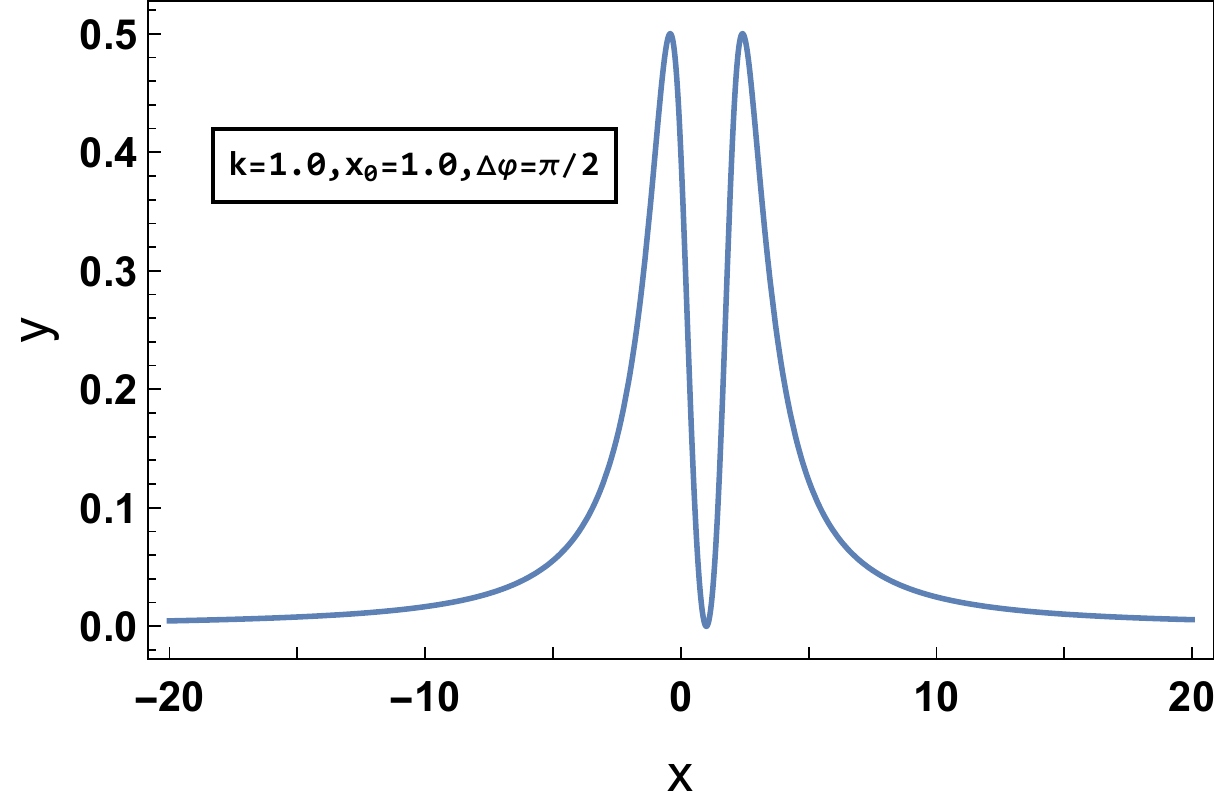}}
     \centerline{(c)}
\end{minipage}
\hfill
\begin{minipage}{0.45\linewidth}
	\centerline{\includegraphics[width=7.0cm]{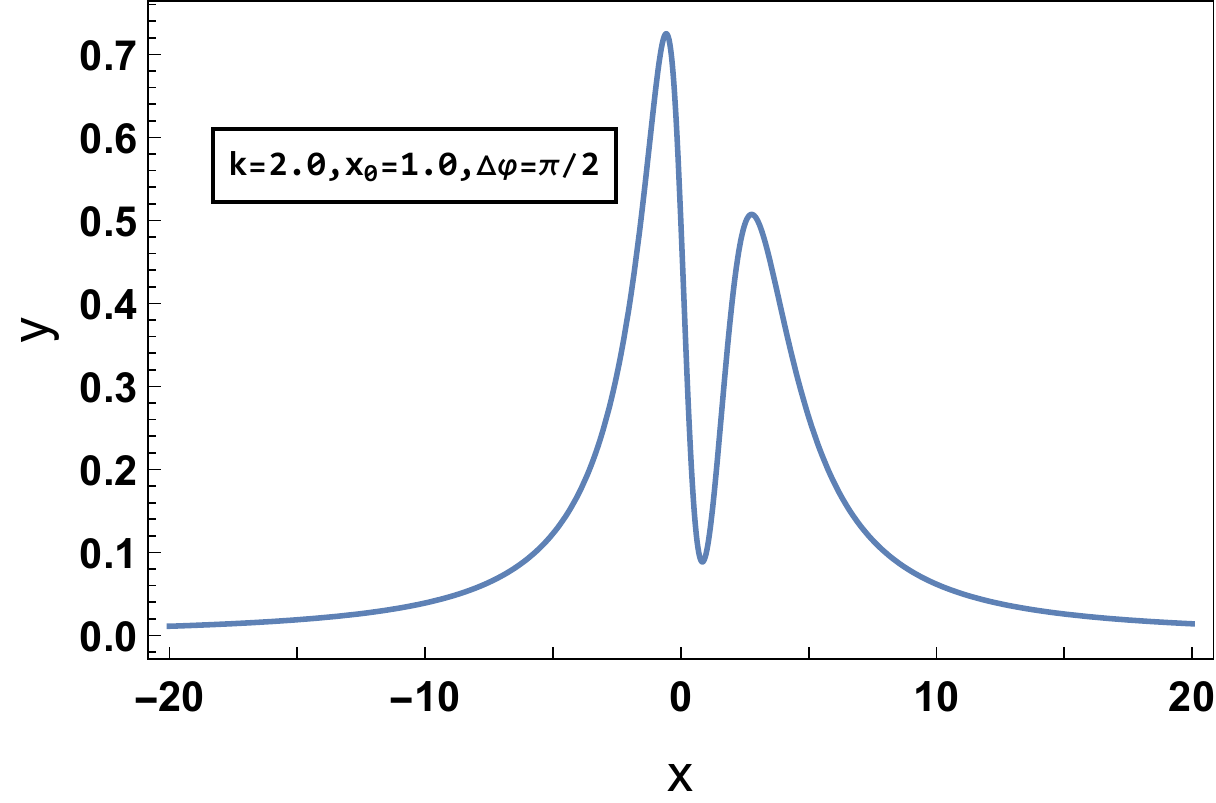}}
     \centerline{(d)}
\end{minipage}
\hfill
\caption{$y$-function in the two-generations case.}
\label{2geny}
\end{figure}

We first consider a simple case. That is, we assume $|U_{\ell 5}|$ and $|U_{\ell 4}|$ are flavor universal. From Eq.~(2) and Eq.~(3) we get $k=k_\ell$ and $k_{\mu e}=k_{\tau e}=1$, which indicates that $k$ does not depend on the neutrino mass. As a result, the functions $y$, Iy, and $R_{21}$ will depend only on $k$, $x_0$, and $\Delta\varphi$.  We display our result of the $y$-function in Fig.~\ref{2geny}, with different choices of the parameter $(k,x_0,\Delta\varphi)$. From Fig.~3(b) and (c) we can see $\Delta\varphi$ affects the shape and hight of the peak. From Fig.~3(a) and (c) we can get a similar message. This means as $x_0$ or $\Delta\varphi$ changes, the interference effect between two generations of sterile neutrinos also changes. Comparing Fig.~3(c) and (d), we can see the deviation from $k=1$ will result in the discrepancy of the heights of two peaks. If $k$ gets either too small or too large, only $N_4$ or $N_5$ give the main contribution, and this returns to the one-generation case.

The function Iy in this situation has the following form
\begin{equation}\label{Iysimple}
\text{Iy}(k,x_0,\Delta\varphi)=\pi(k+1)+\frac{4\pi k}{(k+1)^2+4x_0^2}[(k+1)\cos\Delta\varphi-2x_0\sin\Delta\varphi].
\end{equation}
By defining $\cos\eta=2x_0/\sqrt{(k+1)^2+4x_0^2}$ and $\sin\eta=(k+1)/\sqrt{(k+1)^2+4x_0^2}$, we can rewrite Eq.~(13) as
\begin{equation}
\text{Iy}(k,x_0,\Delta\varphi)=\pi(k+1)+\frac{4\pi k}{k+1}\sin\eta\sin(\eta-\Delta\varphi).
\end{equation}
Correspondingly, the function $R_{21}$ becomes
\begin{equation}
R_{21}(k,x_0,\Delta\varphi)=\left[k+1+\frac{4k}{k+1}\sin\eta\sin(\eta-\Delta\varphi)\right]^{-1/2}.
\end{equation}

In Fig.~4 we show how $R_{21}$ changes with the parameters. From Fig.~1(a) one can see that with $k=1$, there is a peak, whose value and position change with $x_0$. As $x_0$ decreasing, the value of the peak is enhanced and its position moves toward to $\Delta\varphi=\pi$. This means that the difference between two-generation case and one-generation case gets larger when $x_0$ gets smaller if we take $\Delta\varphi$ around $\pi$. From Fig.~1(b), one can see that if we set $\Delta\varphi=\pi$ and change $k$, the value of $R_{21}$ can also be greatly affected. This means if we want a large $R_{21}$, we should take a value of $k$ not far from 1.

\begin{figure}[htbp]
	\begin{minipage}{0.45\linewidth}
		\centerline{\includegraphics[width=7.0cm]{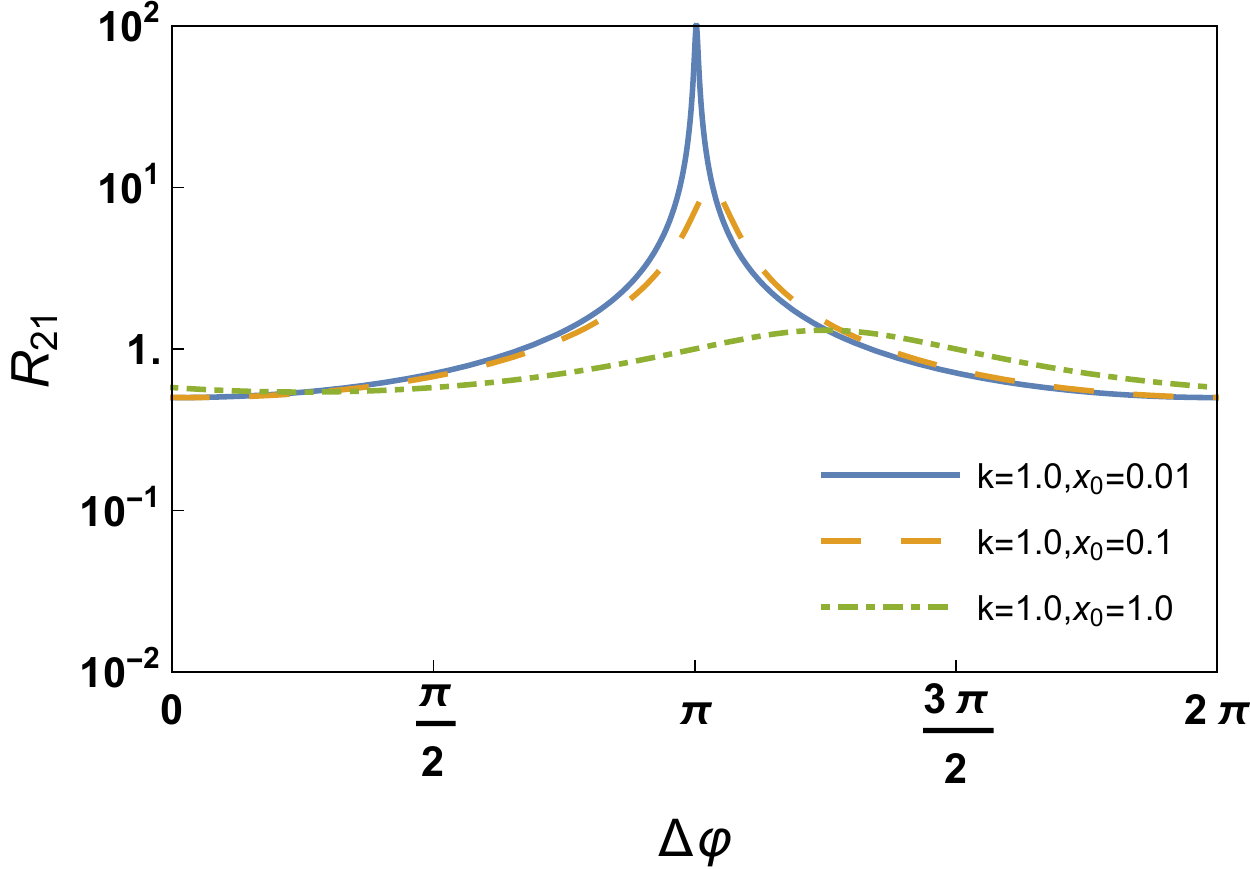}}
		\centerline{(a)}
	\end{minipage}
	\hfill
	\begin{minipage}{0.45\linewidth}
		\centerline{\includegraphics[width=7.0cm]{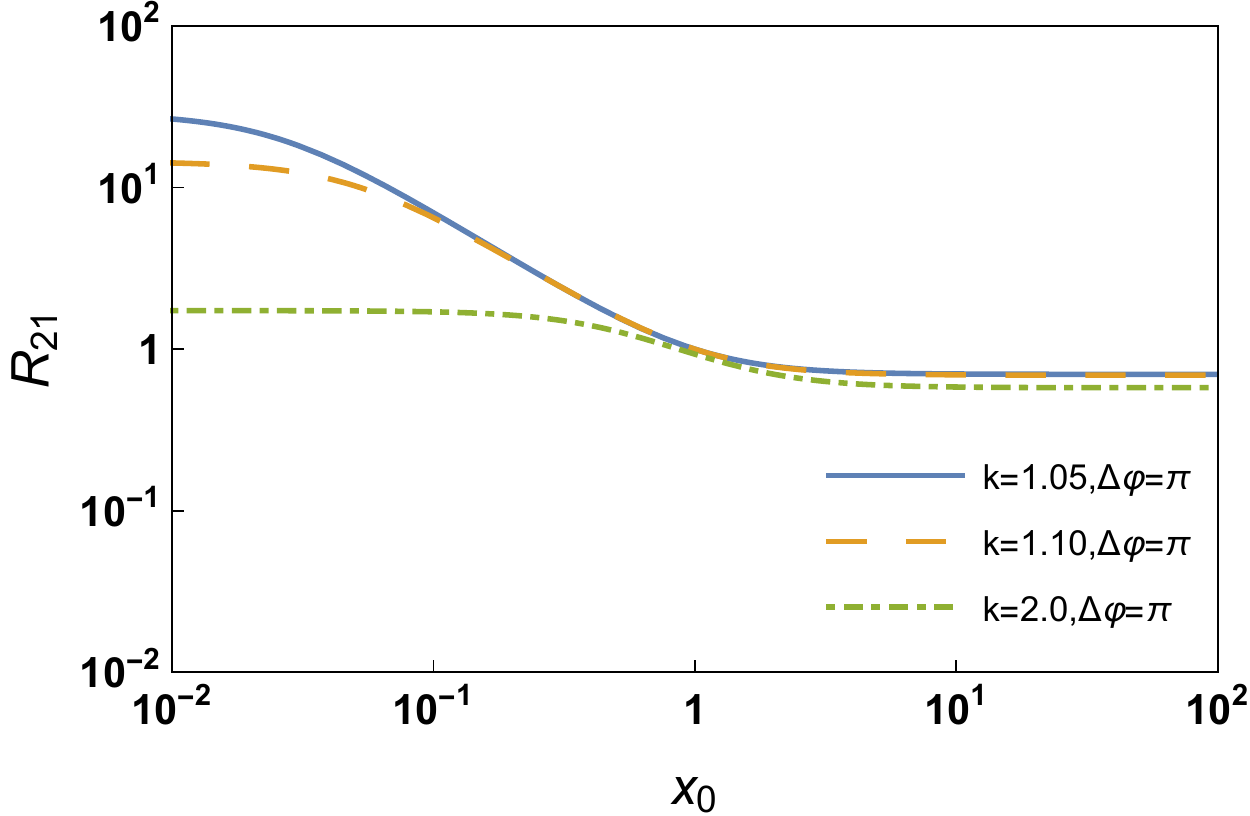}}
		\centerline{(b)}
	\end{minipage}
	\vfill
	\caption{$R_{21}$ as a function of $k$, $x_0$, and $\Delta\varphi$.}
	\label{kneq1}
\end{figure}


Next we consider a situation with the constraint a little relaxed. That is we only assume $k_{\mu e}=k_{\tau e}=1$, but leave the ratio $k_e$, $k_{\mu}$, and  $k_{\tau}$ as free parameters. As a result, $k$ will have the following form
\begin{equation}
k=\frac{k_ef_e(m_4)+k_{\mu}f_\mu(m_4)+k_{\tau}f_\tau(m_4)}{f_e(m_4)+f_\mu(m_4)+f_\tau(m_4)}.
\end{equation}
One can see that generally $k$ will depend both on $k_\ell$ and the neutrino mass. Correspondingly, $R_{21}$ will also depend on such parameters. As an example, in Fig.~5 we present the dependence of $R_{21}$ on $k$ and $k_e$ with $x_0=0.1$ and $\Delta\varphi=\pi$. We can see that $R_{21}$ will reach its maximum when both $k$ and $k_e$ are around one.  From Eq.~(16) we get $k\ge f_ek_e/\sum_{\ell} f_\ell$. Only the ranges above the straight line (with a specific value of $m_4$) in Fig.~5 are allowed.
\begin{figure}[htbp]
\centering
\includegraphics[width=7.0cm]{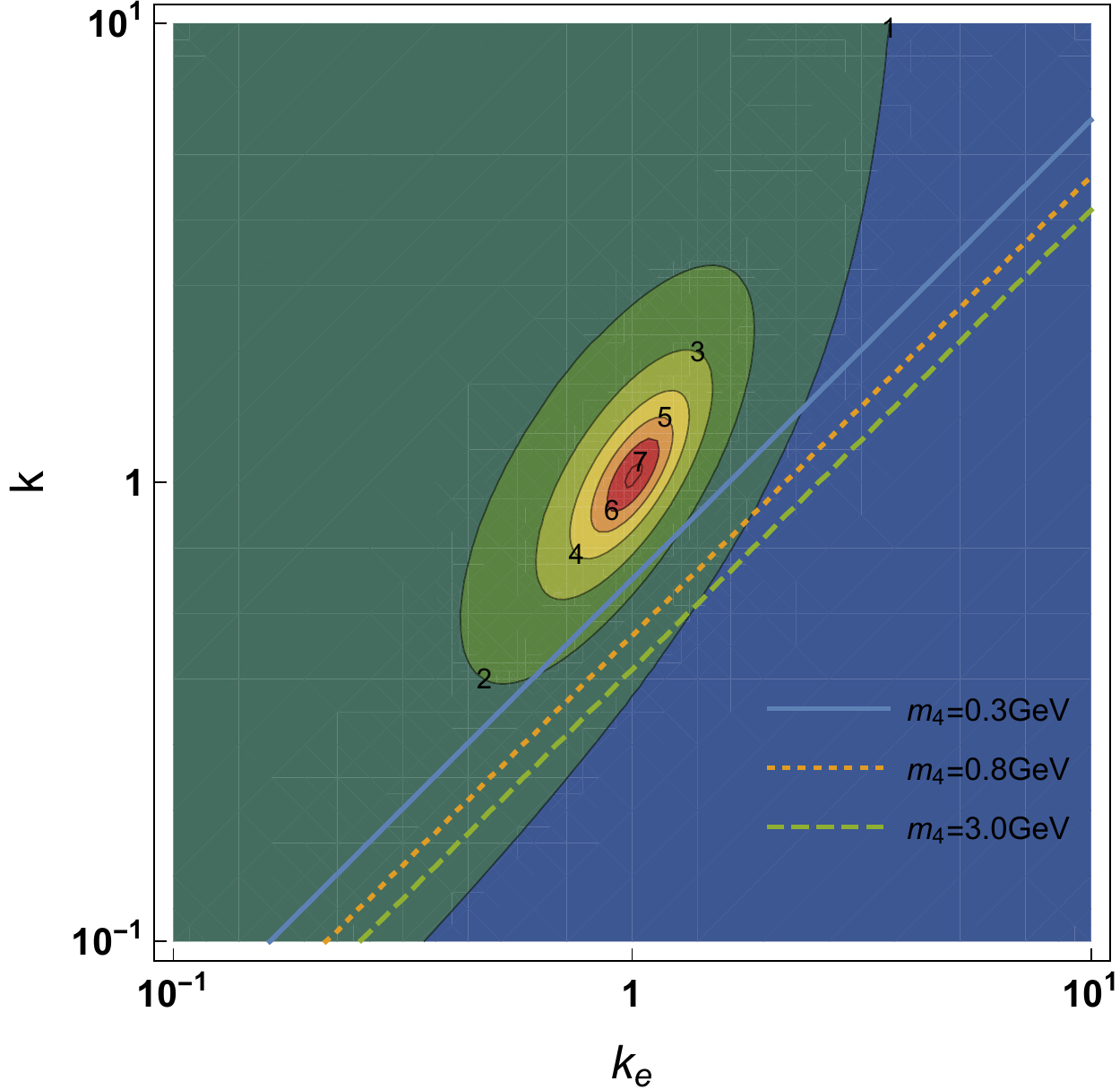}
\caption{The dependence of $R_{21}$ on $k$ and $k_e$ with $x_0=0.1$ and $\Delta\varphi=\pi$.}
\label{kek1}
\end{figure}

The dependence of $R_{21}$ on $m_4$ are presented in Fig.~6, where specific values of the parameters are assumed. One notices that $R_{21}$ strongly depends on the sterile neutrino mass when the later is less than 1 GeV (except the $k_e=k_\mu=k_\tau=1$ case).  When $m_4$ is larger than 1 GeV, the curves become smooth. The reason for this is that $f_e(m_4)$,  $f_\mu(m_4)$, and $f_\tau(m_4)$ have a similar dependence on $m_4$ (see Fig.~1(a)), and $R_{21}$ depend on these functions only through $k$ given in Eq.~(16). Then we study how these results will affect the upper limits of the mixing parameters. By using the experimental results of the LNV processes of $K$ and $D$ mesons in Table I, together with Fig.~6 and Eq.~(11), we get the upper limits of the mixing parameters with different sterile neutrino masses. The results, which are represented as the solid lines,  are presented in Fig.~7. For comparison, we also presented the results of one-generation case (the dotdashed lines). We can see that with such a choice of parameters, the square of the mixing parameters could be raised approximately by two orders of magnitude. 

\begin{figure}[htbp]
\begin{minipage}{0.45\linewidth}
	\centerline{\includegraphics[width=6.5cm]{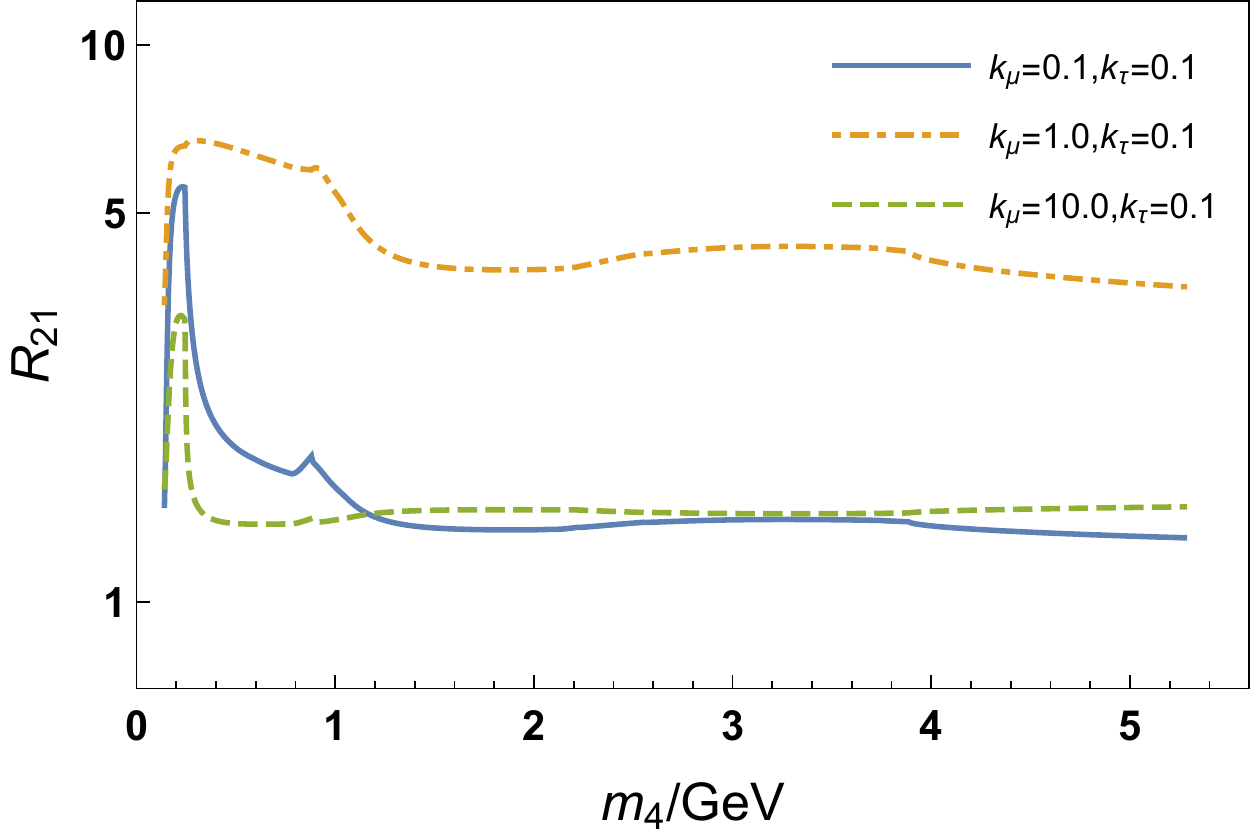}}
     \centerline{(a)}
\end{minipage}
\hfill
\begin{minipage}{0.45\linewidth}
	\centerline{\includegraphics[width=6.5cm]{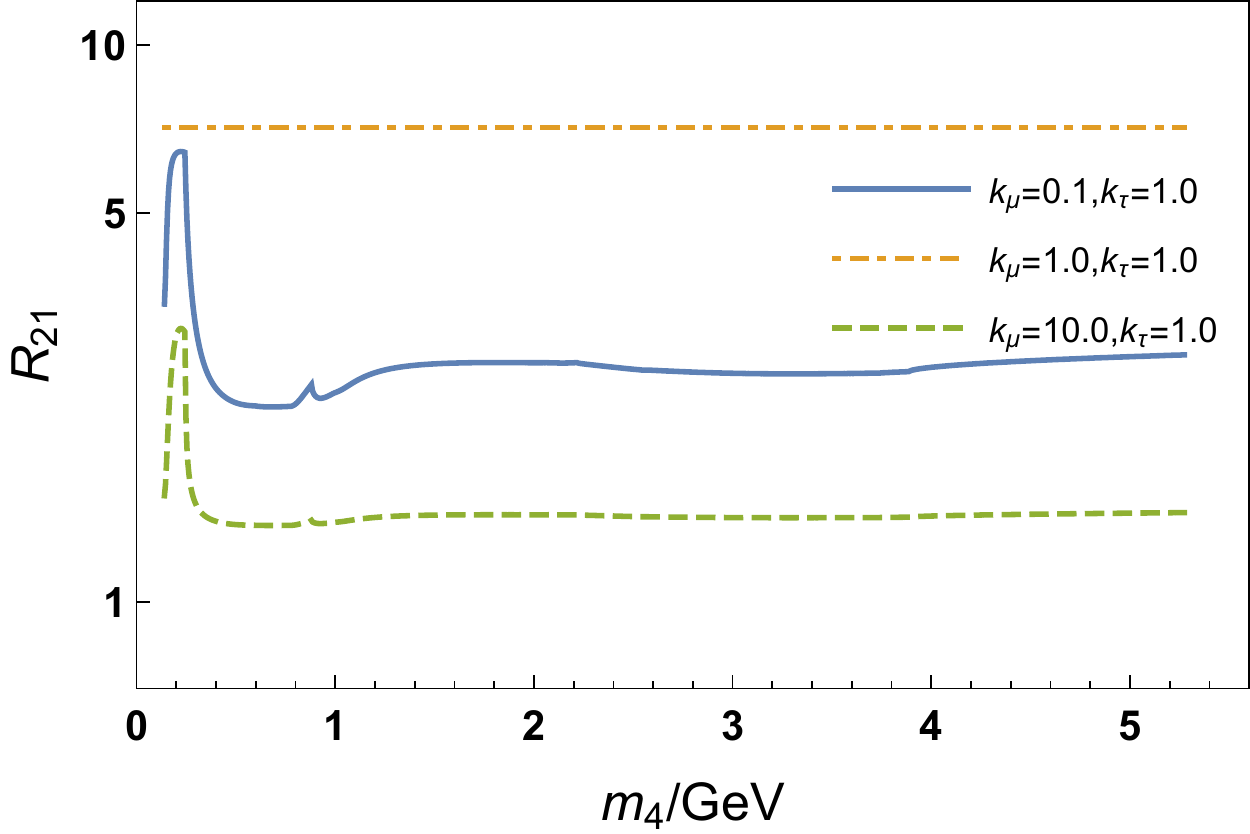}}
     \centerline{(b)}
\end{minipage}
\vfill
\begin{minipage}{0.45\linewidth}
	\centerline{\includegraphics[width=6.5cm]{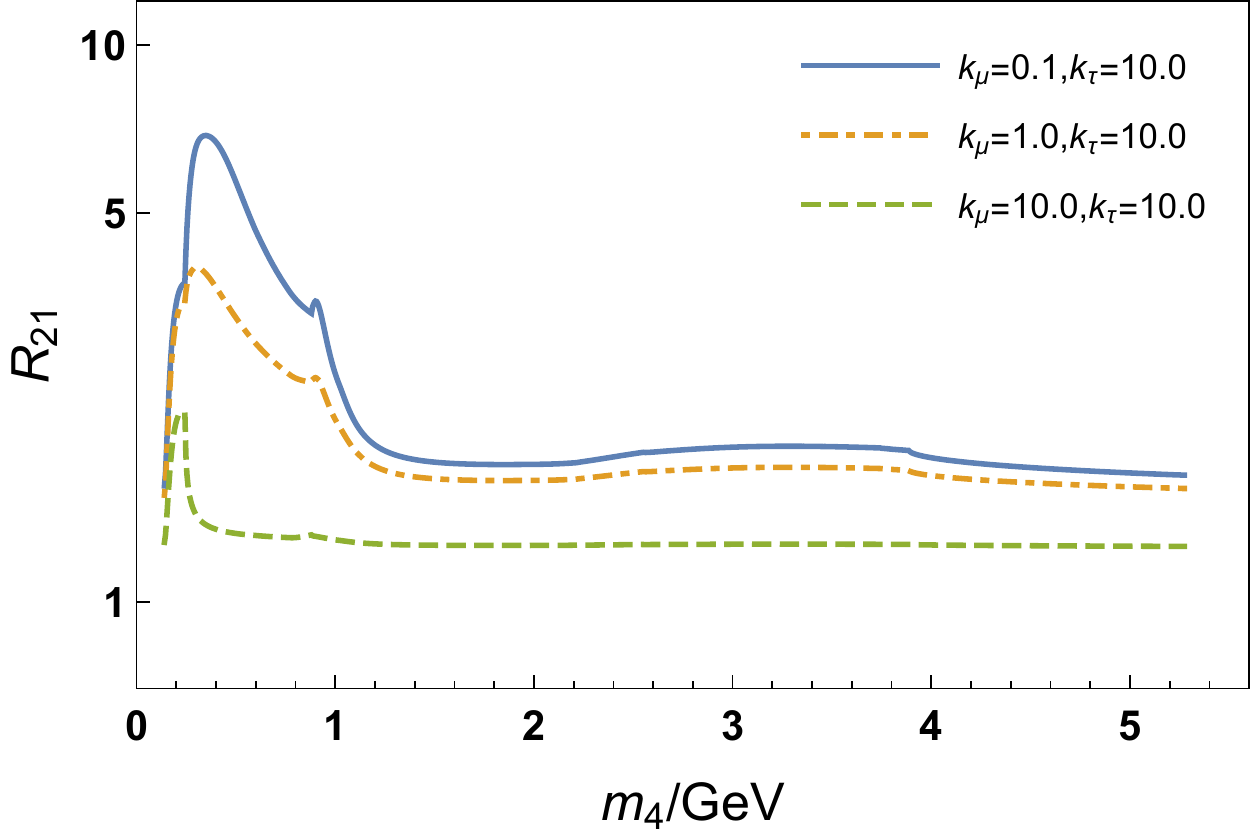}}
     \centerline{(c)}
\end{minipage}
\hfill
\caption{The dependence of $R_{21}$ on $m_4$ with different values of $k_\mu$ and $k_\tau$. Here we have chosen $k_e=k_{\mu e}=k_{\tau e}=1$, $x_0=0.1$, and $\Delta\varphi=\pi$.}
\label{R}
\end{figure}

\begin{figure}[htbp]
\begin{minipage}{0.45\linewidth}
	\centerline{\includegraphics[width=6.5cm]{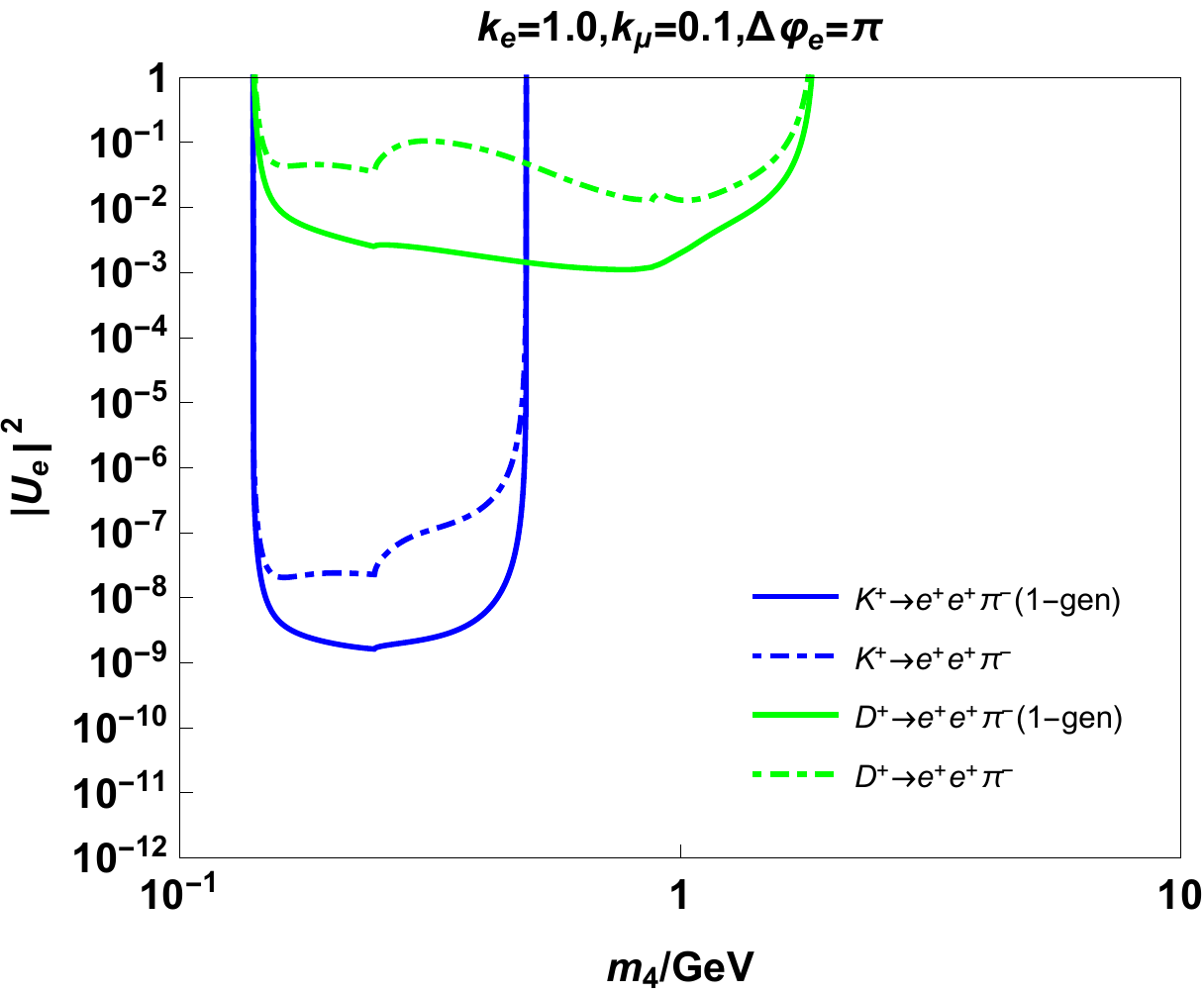}}
     \centerline{(a)}
\end{minipage}
\hfill
\begin{minipage}{0.45\linewidth}
	\centerline{\includegraphics[width=6.5cm]{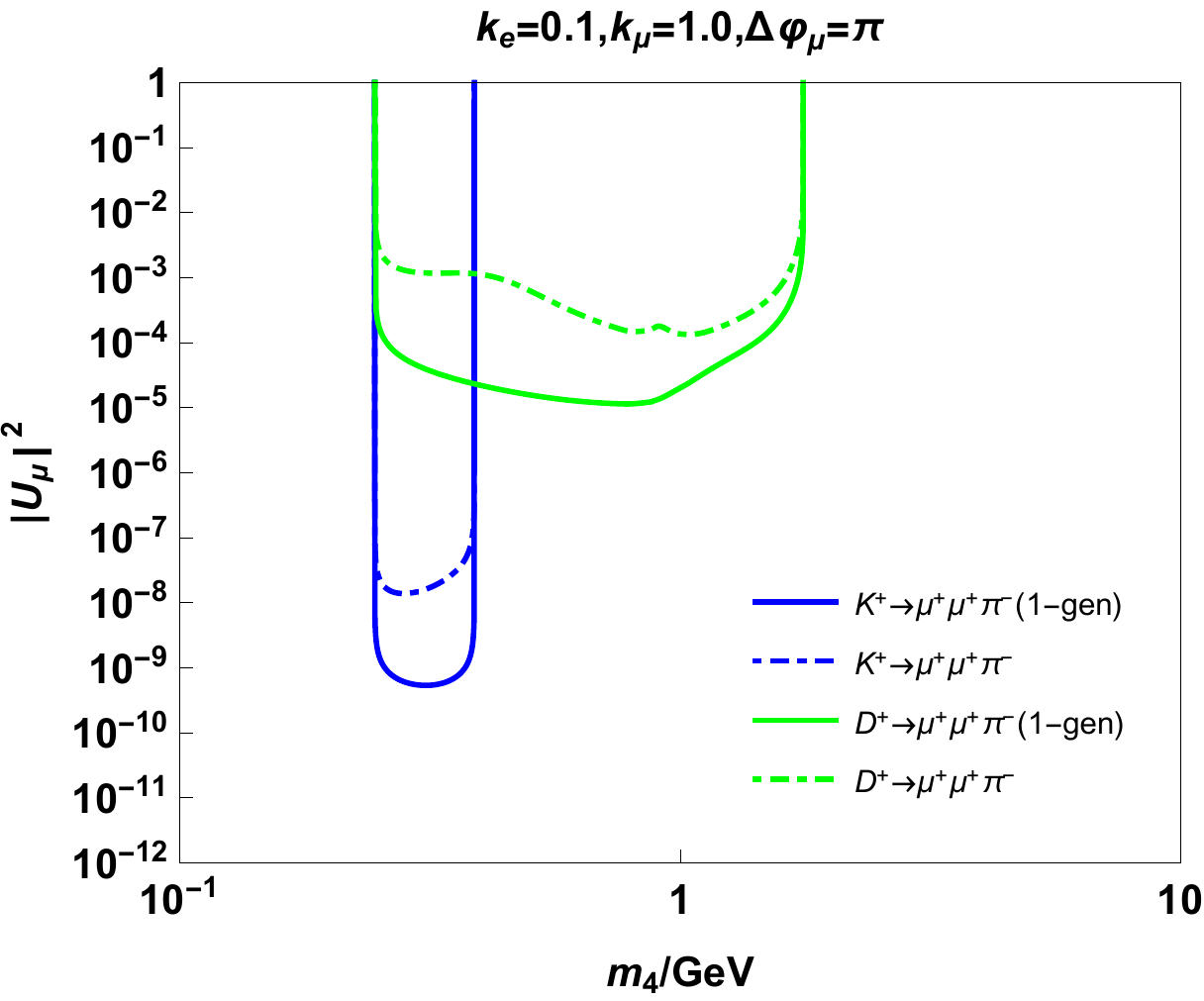}}
     \centerline{(b)}
\end{minipage}
\hfill
\begin{minipage}{0.45\linewidth}
	\centerline{\includegraphics[width=6.5cm]{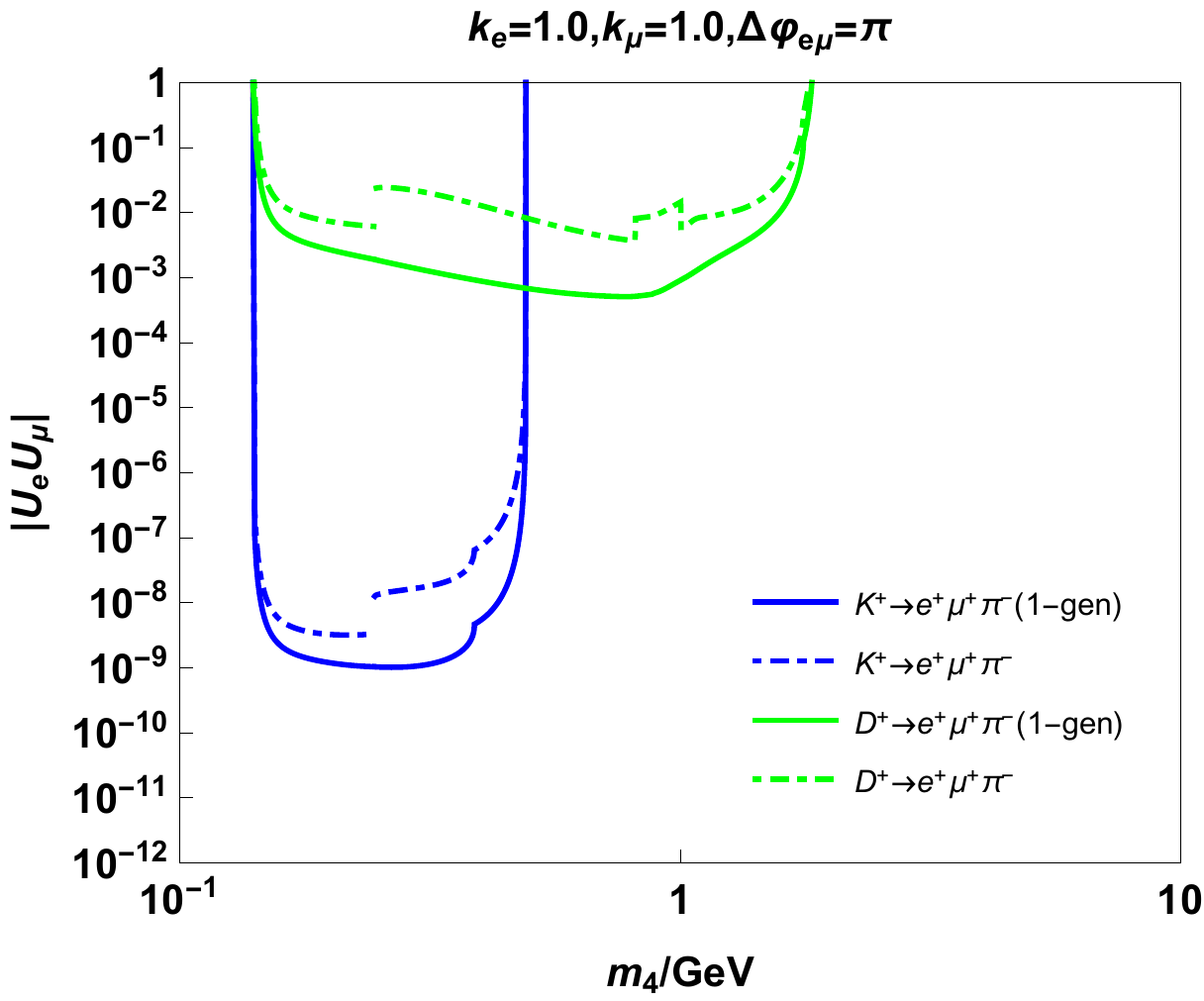}}
     \centerline{(c)}
\end{minipage}
\hfill
\caption{Experimental bounds on $|U_{\ell_1}U_{\ell_2}|$ in one-generation and two-generation scenarios. For the later case, we choose $k_{\mu e}=k_{\tau e}=1$, $k_{\tau}=10.0$ and $x_0=0.1$.}
\label{Ul}
\end{figure}

\begin{table}[htb]
	\setlength{\tabcolsep}{0.2cm}
	\caption{Experimental results of the LNV processes of $K^+$ and $D^+$ mesons~\cite{Tanabashi:2018oca}.}
	\label{EULUe}
	\centering
	\begin{tabular*}{\textwidth}{c @{\extracolsep{\fill}} ccc}
		\hline\hline
		Decay channel & Braching Ratio &Decay channel & Braching Ratio \\
		\hline
		$K^+\rightarrow e^+e^+ \pi^-$& $<6.4\times10^{-10}$ & $D^+\rightarrow e^+e^+ \pi^-$ & $<1.1\times10^{-6}$\\
		\hline
		$K^+\rightarrow \mu^+\mu^+ \pi^-$& $<8.6\times10^{-11}$ & $D^+\rightarrow\mu^+\mu^+ \pi^-$ & $<2.2\times10^{-8}$ \\	
		\hline
		$K^+\rightarrow e^+\mu^+ \pi^-$& $<5.0\times10^{-10}$ & $D^+\rightarrow e^+\mu^+ \pi^-$& $<2.0\times10^{-6}$\\
		\hline\hline
	\end{tabular*}
\end{table}

\section{CP violation}\label{CP}

If there is only one generation of heavy sterile neutrino, no CP violation in such LNV processes will be generated. So to study CP asymmetry, we should consider at least two generations of heavy sterile neutrinos. We first consider the LNV processes $K^{\pm}\rightarrow e^{\pm}e^{\pm}\pi^{\mp}$. The CP asymmetry of such decay channels is defined as
\begin{equation}
\mathcal{A}_{CP}=\frac{\Gamma(K^-\rightarrow e^-e^-\pi^+)-\Gamma(K^+\rightarrow e^+e^+\pi^-)}{\Gamma(K^-\rightarrow e^-e^-\pi^+)+\Gamma(K^+\rightarrow e^+e^+\pi^-)}.
\end{equation}
It turns out that the only difference between the decay widths of these two CP-conjugated channels lies in the phases of the active-sterile mixing parameters. Specifically, we just need to change the sign of the CP phase $\Delta\varphi$ to get $\Gamma(K^-\rightarrow e^-e^-\pi^+)$ from $\Gamma(K^+\rightarrow e^+e^+\pi^-)$. From Eq.~(9) and Eq.~(10), we get
\begin{equation}\label{ACPLNV}
\begin{aligned}
\mathcal{A}_{CP}&=\frac{\text{Iy}(k_e,k,x_0,-\Delta\varphi)-\text{Iy}(k_e,k,x_0,\Delta\varphi)}{\text{Iy}(k_e,k,x_0,-\Delta\varphi)+\text{Iy}(k_e,k,x_0,\Delta\varphi)}\\
&=\frac{8kx_0\sin\Delta\varphi}{(k_e+k/k_e)[(k+1)^2+4x_0^2]+4k(k+1)\cos\Delta\varphi}.
\end{aligned}
\end{equation}
As $k$ generally depends on $k_\ell$, $k_{\mu e}$, $k_{\tau e}$, and the $m_4$, the CP asymmetry will also depend on these parameters.

For the decay processes $K^{\pm}\rightarrow e^{\pm}\mu^{\pm}\pi^{\mp}$, the function Iy will depend on $k_\mu$ both directly and indirectly. In this case, the CP asymmetry can be written as
\begin{equation}\label{ACPLFV}
\begin{aligned}
\mathcal{A}_{CP}&=\frac{\text{Iy}(k_e,k_{\mu},k,x_0,-\Delta\varphi)-\text{Iy}(k_e,k_{\mu},k,x_0,\Delta\varphi)}{\text{Iy}(k_e,k_{\mu},k,x_0,-\Delta\varphi)+\text{Iy}(k_e,k_{\mu},k,x_0,\Delta\varphi)}\\
&=\frac{8kx_0\sin\Delta\varphi}{(\sqrt{k_ek_{\mu}}+k/\sqrt{k_ek_{\mu}})[(k+1)^2+4x_0^2]+4k(k+1)\cos\Delta\varphi},
\end{aligned}
\end{equation}
where $\Delta\varphi=(\phi_{e5}-\phi_{e4})+(\phi_{\mu5}-\phi_{\mu4})$. One can see Eq.~(19) can be achieved from Eq.~(18) by replacing $k_e$ by $\sqrt{k_ek_\mu}$.


As a first step, we consider the simple situation. That is, $|U_{\ell4}|$ and $|U_{\ell5}|$ are flavor-universal, from which we get $k=k_\ell$ and $k_{\mu e}=k_{\tau e}=1$. Then Eq.~(19) can be simplified as
\begin{equation}\label{RCP}
\mathcal{A}_{CP}=\frac{8kx_0\sin\Delta\varphi}{(k+1)[(k+1)^2+4x_0^2+4k\cos\Delta\varphi]}.
\end{equation}
One notices that this result is independent of the initial meson and the sterile neutrino mass. From Eq.~\eqref{RCP} we can see $\mathcal{A}_{CP}(\pi+\Delta\varphi)=-\mathcal{A}_{CP}(\pi-\Delta\varphi)$, so we will only focus on the region $\Delta\varphi\in(0,\pi)$. In Fig.~8, we present the numerical results of $\mathcal{A}_{CP}$ changing with $\Delta\varphi$ and $x_0$. Here we have chosen $k=1$ for simplicity. One can see $\mathcal{A}_{CP}$ has maximum when $x_0\rightarrow0$ and $\Delta\varphi\rightarrow\pi$.

\begin{figure}[htbp]
\begin{minipage}{0.45\linewidth}
	\centerline{\includegraphics[width=6.0cm]{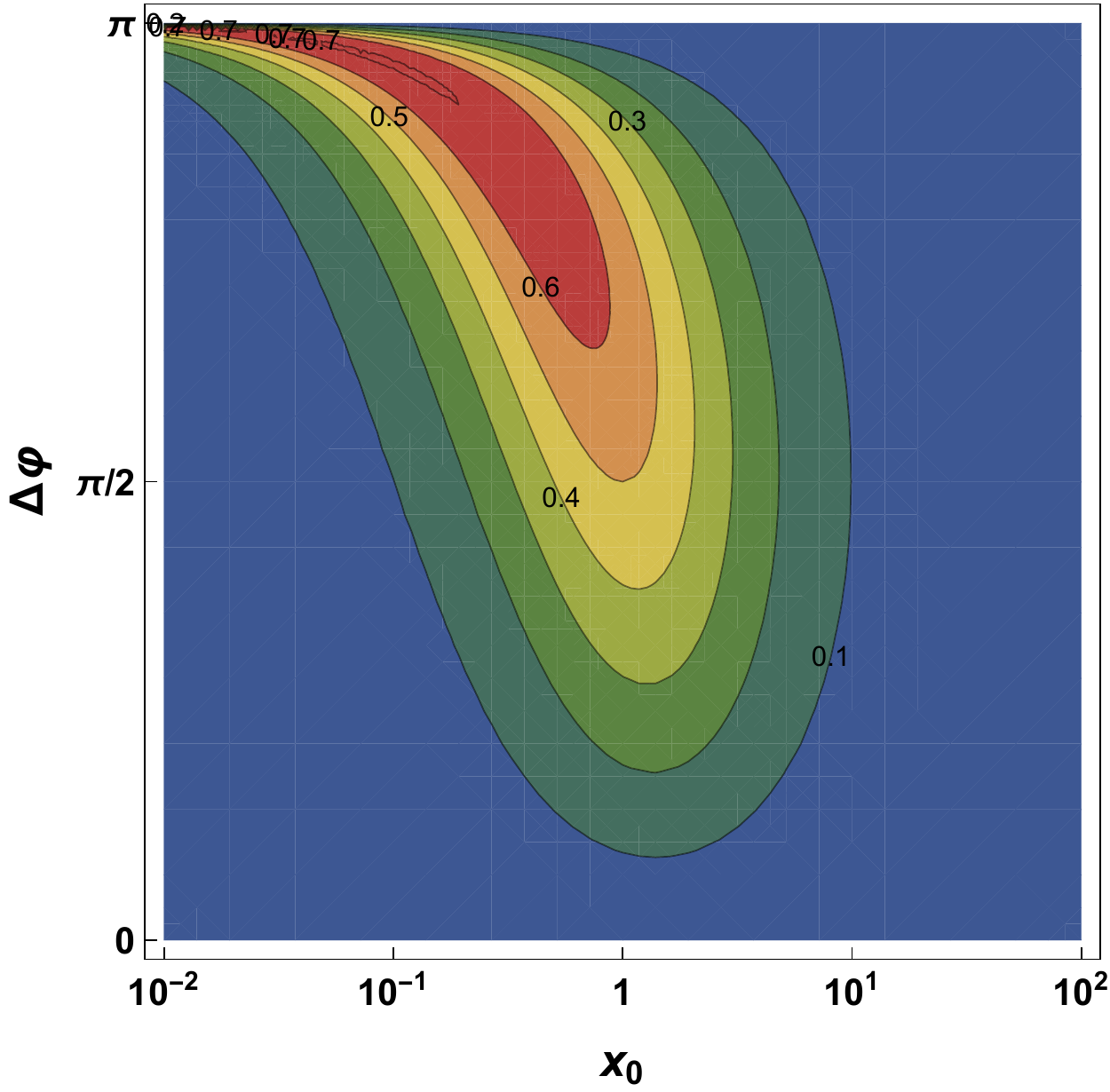}}
     \centerline{(a)}
\end{minipage}
\hfill
\begin{minipage}{0.45\linewidth}
	\centerline{\includegraphics[width=6.2cm]{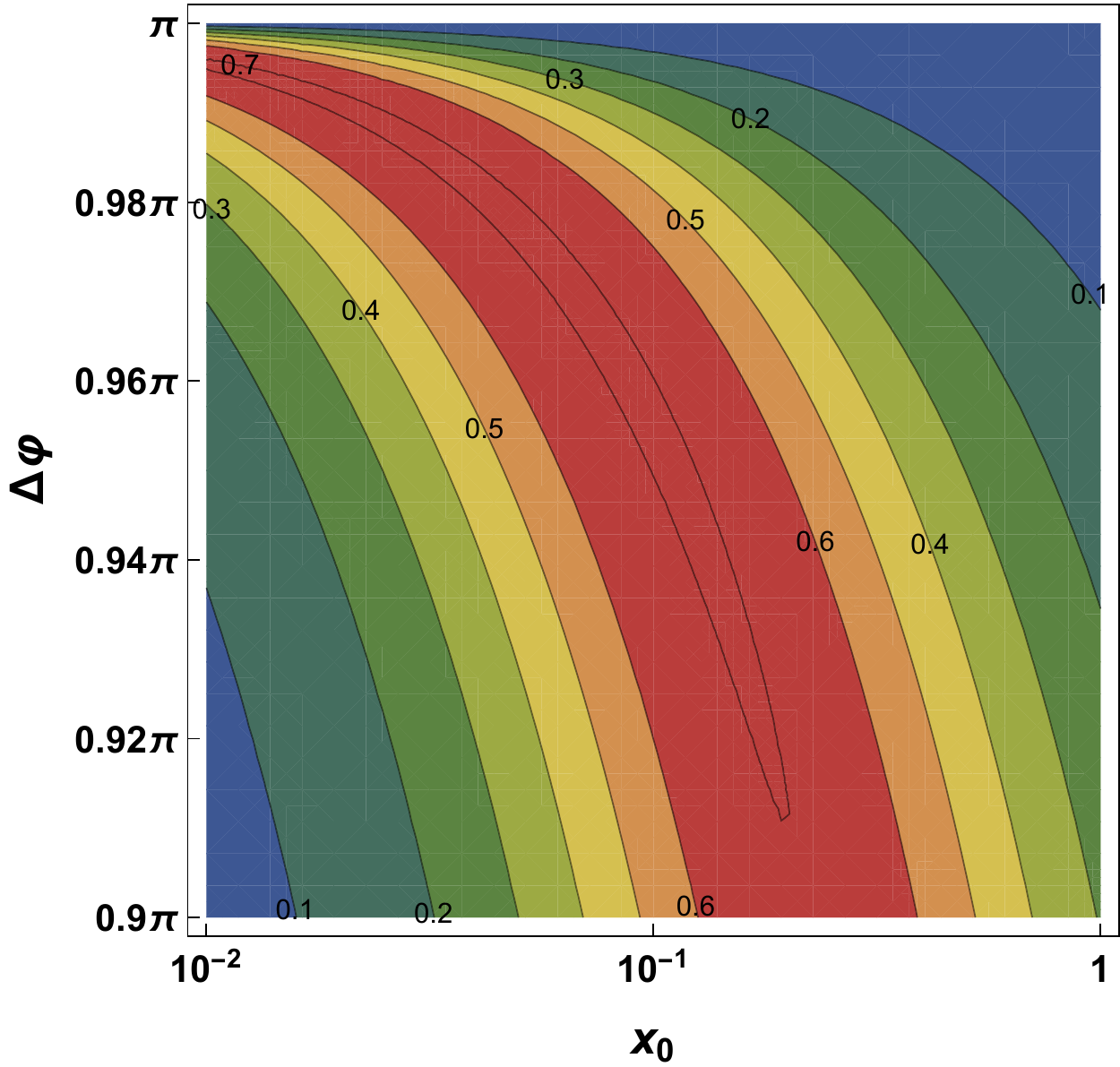}}
     \centerline{(b)}
\end{minipage}
\vfill
\caption{(a)~$\mathcal{A}_{CP}$ as a function of $x_0$ and $\Delta\varphi$ with $k=1$; (b)~the region around the maximum.}
\label{ACP1}
\end{figure}

It is interesting to find the maximum of $\mathcal{A}_{CP}$. To that end, we define $\alpha=\pi-\Delta\varphi$, and from the discussion above we know the maximum can be achieved only when $\alpha$ is small. Then we get
\begin{equation}
\begin{aligned}
\mathcal{A}_{CP}(k=1,x_0,\Delta\varphi=\pi-\alpha)&=\frac{x_0\sin\alpha}{x_0^2+1-\cos\alpha}\\
&\approx\frac{x_0\alpha}{x_0^2+\alpha^2/2}=\frac{\beta}{1/2+\beta^2},
\end{aligned}
\end{equation}
where $\beta\equiv x_0/\alpha$. One can see that in this special case, the CP asymmetry only depends on the ratio of $x_0$ and $\alpha$, namely $\beta$. The maximum of $\mathcal{A}_{CP}$ is $\sqrt{2}/2$, which can be achieved when $\beta=\sqrt{2}/2$.


Next we consider the case with a relaxed condition, that is $k_{\mu e}=k_{\tau e}=1$, but leave $k_e$, $k_\mu$, and $k_\tau$ as free parameters.     $\mathcal{A}_{CP}$ will generally depend both on $k_l$ and $m_4$. We want to find its extremum value under such condition. To that end, we take the partial derivatives of $\mathcal{A}_{CP}$ with respect to $x_0$ and $\Delta\varphi$, and set these derivatives to zero,
\begin{equation}
\frac{\partial}{\partial x_0}\mathcal{A}_{CP}=0,~~~\frac{\partial}{\partial\Delta\varphi}\mathcal{A}_{CP}=0.
\end{equation}
By solving these equations we get the extreme point,
\begin{equation}
4x_0^2|_{\text{extre}}=\sqrt{(k_e+1)^4-16k_e^2},~~~\cos\Delta\varphi|_{\text{extre}}=\frac{4x_0^2|_{\text{extre}}-(k_e+1)^2}{4k_e}.
\end{equation}
Submitting them into Eq.~\eqref{ACPLNV}, we obtain
\begin{equation}
\mathcal{A}_{CP_{\text{extre}}}(k_e,k)=\frac{8 k x_0^2|_{\text{extre}}\sqrt{-2k_e\cos\Delta\varphi|_{\text{extre}}}}{4x_0^2|_{\text{extre}}(k^2+2 k+k_e^2)+(k+1) (k-k_e)^2}.
\end{equation}

Here we have expressed the extremum of $\mathcal{A}_{CP}$ as the function of $k_e$ and $k$. If we take $k=k_e$ first, and then let $k$ approach to one, we get $\mathcal{A}_{CP_{\text{extre}}}=\sqrt{2}/2$, which is the result in the former case. In Fig.~9, the numerical results are presented. We can see that $\mathcal{A}_{CP_{\text{extre}}}$ is suppressed when $k_e$ or $k$ is either too big or too small. Here only the regions above the straight lines are allowed because of the constraint condition $k\ge f_ek_e/\sum_{\ell} f_\ell$. In general, $k$ depends on $m_4$ (see Eq.~(2)), so does $\mathcal{A}_{CP_{\text{extre}}}$, which is shown in Fig.~10. One can see the parameters $k_l$ can greatly affect the results. We should point out that although our discussion about $\mathcal{A}_{CP}$ is for LNV processes, these results also applies to the LFV processes with a redefinition of $\Delta\varphi$. 

\begin{figure}[htb]
	\centering
	\includegraphics[width=6.5cm]{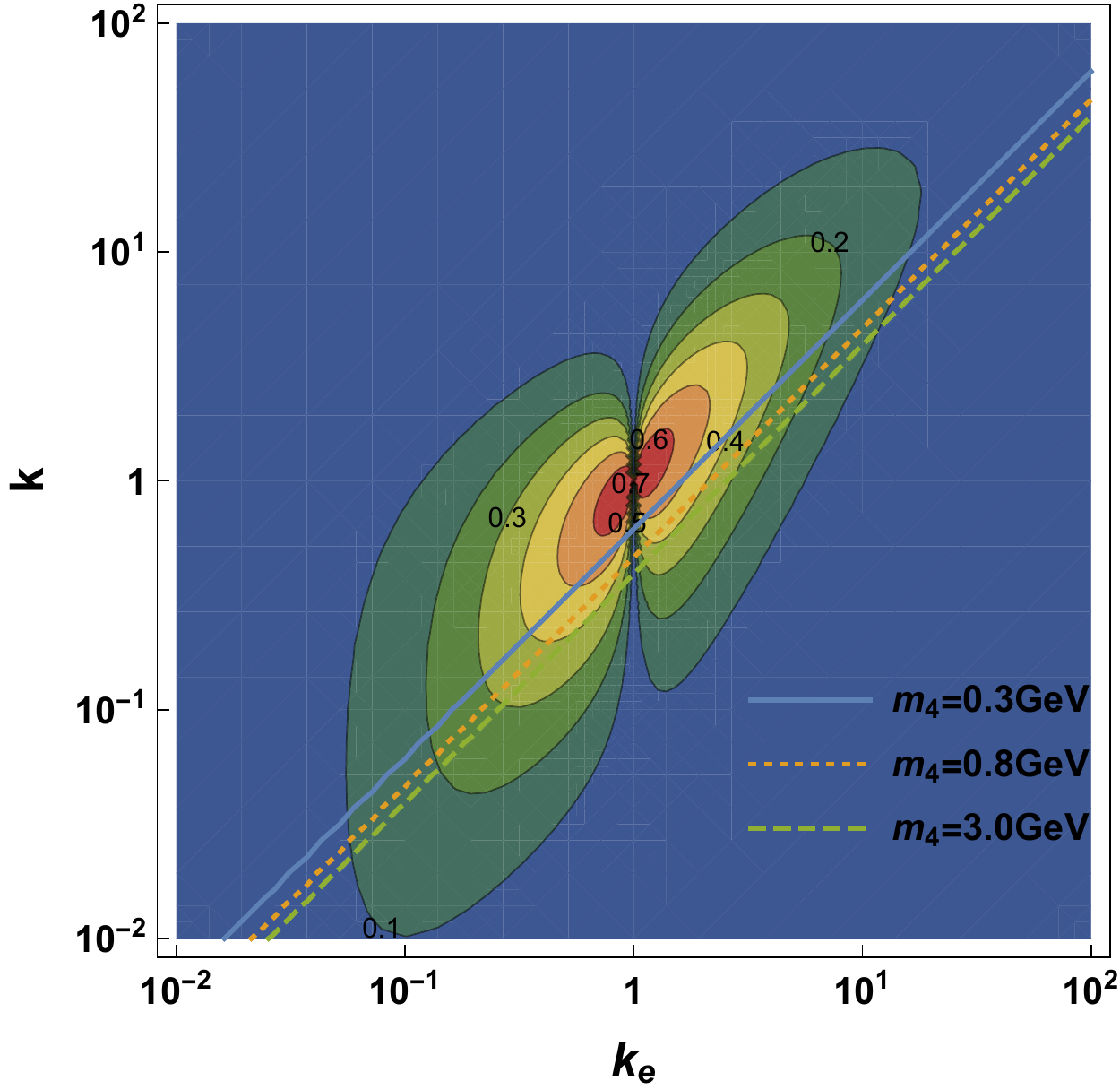}
	\caption{The extremum of $\mathcal{A}_{CP}$ as a function of $k_e$ and $k$.}
	\label{ACPmaxkke}
\end{figure}

\begin{figure}[htb]
	\begin{minipage}{0.45\linewidth}
		\centerline{\includegraphics[width=7.0cm]{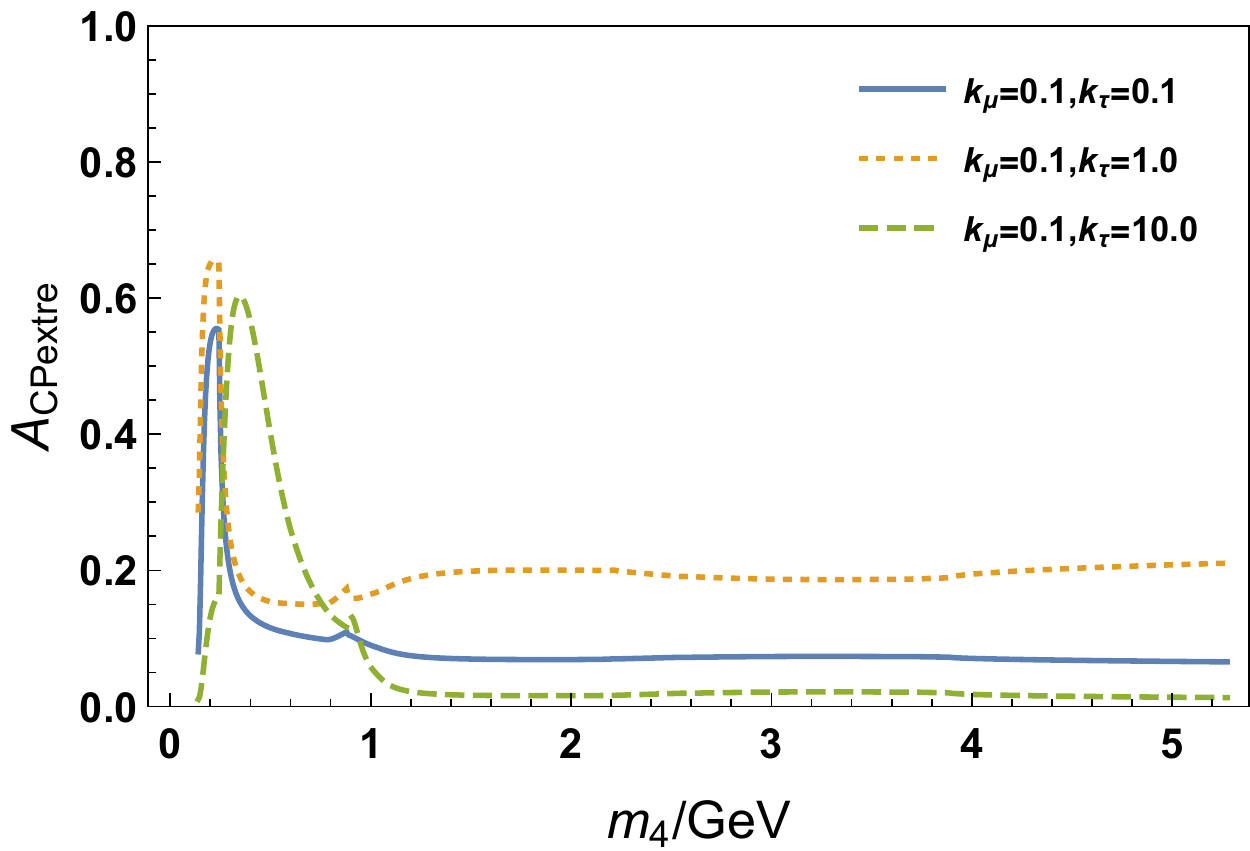}}
		\centerline{(a)$K^{\pm}\rightarrow e^{\pm}e^{\pm}\pi^{\mp}$}
	\end{minipage}
	\hfill
	\begin{minipage}{0.45\linewidth}
		\centerline{\includegraphics[width=7.0cm]{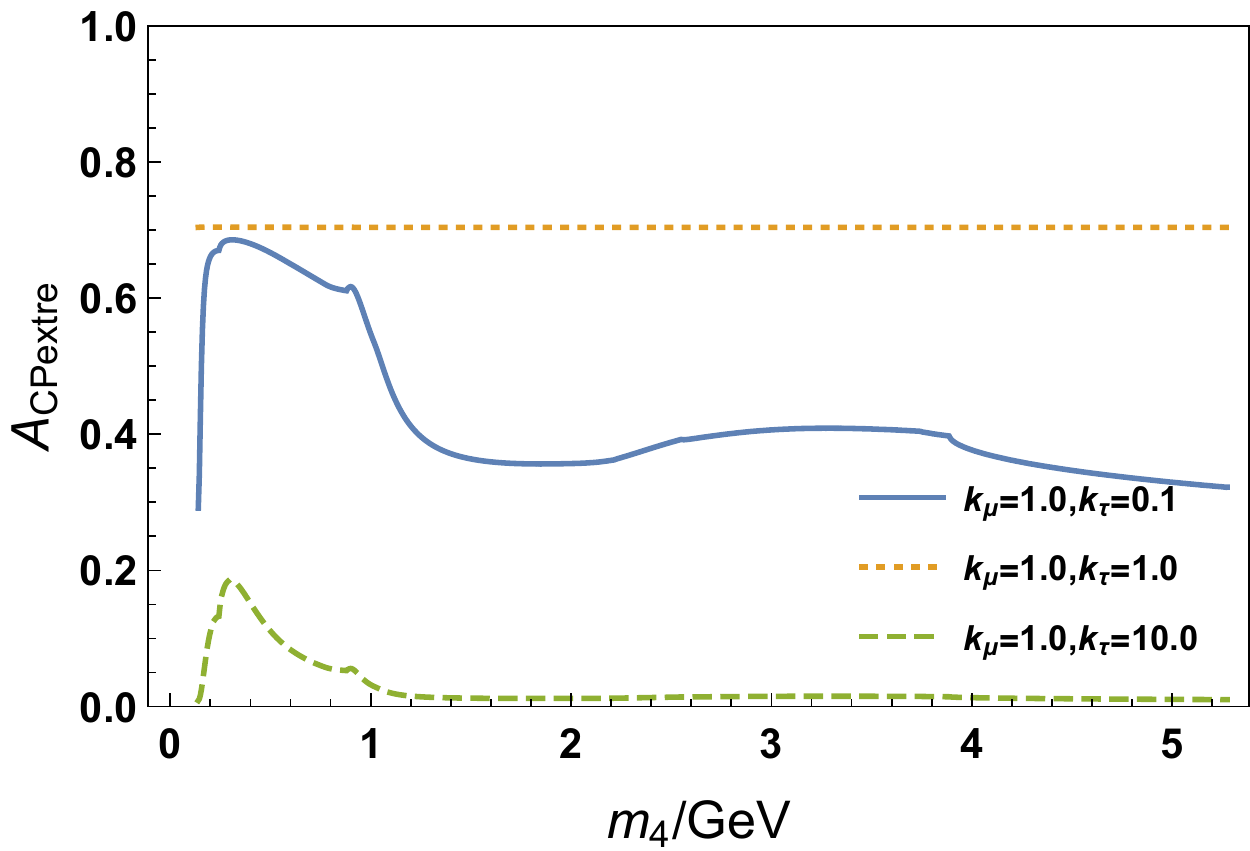}}
		\centerline{(b)$K^{\pm}\rightarrow e^{\pm}e^{\pm}\pi^{\mp}$}
	\end{minipage}
	\vfill
	\begin{minipage}{0.45\linewidth}
		\centerline{\includegraphics[width=7.0cm]{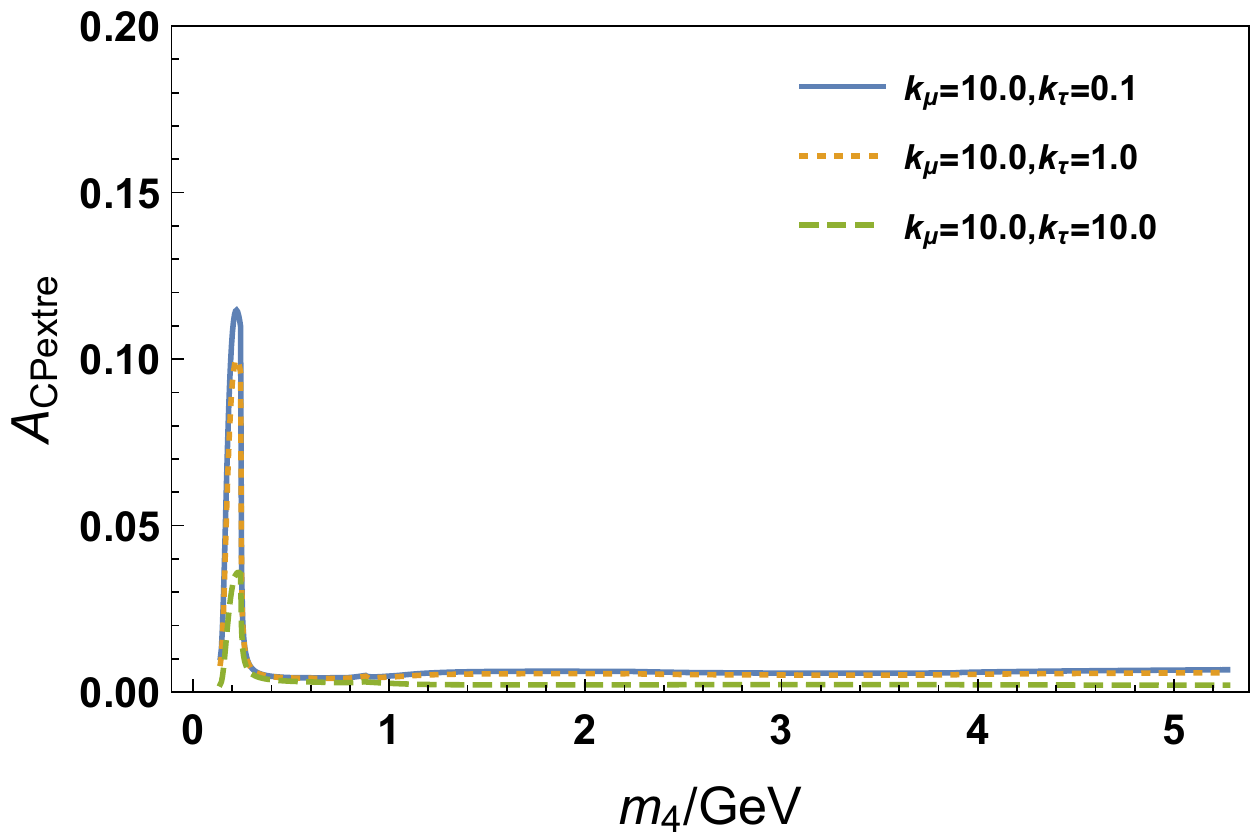}}
		\centerline{(c)$K^{\pm}\rightarrow e^{\pm}e^{\pm}\pi^{\mp}$}
	\end{minipage}
	\hfill
	\begin{minipage}{0.45\linewidth}
		\centerline{\includegraphics[width=7.0cm]{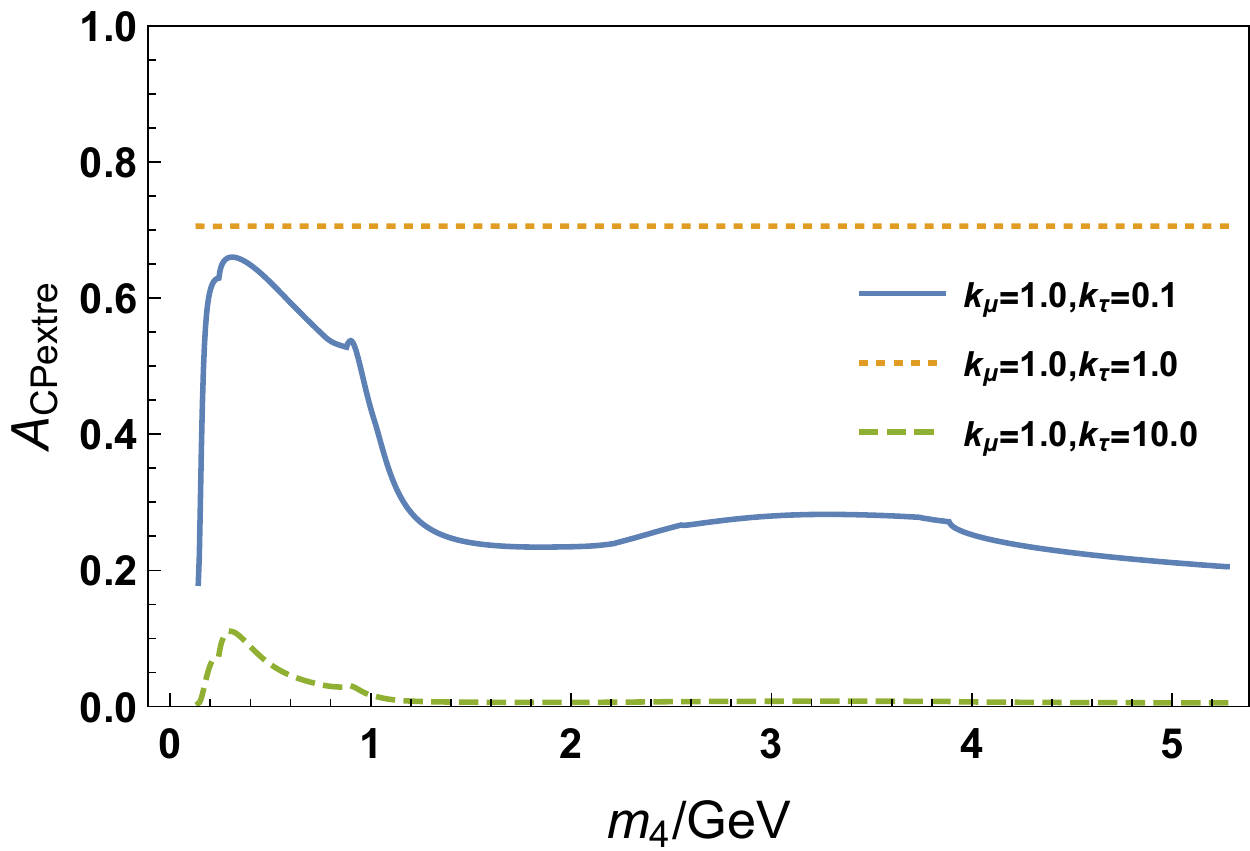}}
		\centerline{(d)$K^{\pm}\rightarrow e^{\pm}\mu^{\pm}\pi^{\mp}$}
	\end{minipage}
	\vfill
	\caption{The mass-denpendecy of the extremum value of CP asymmetry in several choices of parameters, where $k_e=0.99$, $k_{\mu e}=k_{\tau e}=1$.}
	\label{ACPm4}
\end{figure}

\section{Conclusion}

In this paper, we have studied the LNV processes of $K$ and $D$ mesons induced by two quasi-degenerate heavy sterile Majorana neutrinos. Two things are carefully investigated. Firstly, the partial widths of these decays are related to a function Iy, which depends on $k_e$, $k$, $x_0$, and $\Delta\varphi$. Correspondingly, the upper limits of the active-sterile mixing matrix elements extracted by comparing with the experimental data also depend on such parameters. It is shown that when we set $k=1$, $x_0\to 0$, and $\Delta\varphi\to\pi$, there is a big deviation of the results of the two-generation and one-generation cases. Secondly, a general expression for the CP asymmetry of such decay channels is presented. The extremum value of $\mathcal{A}_{CP}$, as a function of $k_e$ and $k$, reaches its maximum value $\sqrt{2}/2$ when we take $k=k_e=1$. Indirectly through $k$, the sterile neutrino mass can greatly affect the extremum value of $\mathcal{A}_{CP}$.

\section{Acknowledgments}

This work was supported in part by the National Natural Science Foundation of China (NSFC) under Grant No.~12075073. We also thank the HEPC Studio at Physics School of Harbin Institute of Technology for access to computing resources through INSPUR-HPC@hepc.hit.edu.cn

\bibliography{main}
\end{document}